\documentclass[11pt,a4paper]{article}
\pdfoutput=1
\usepackage{jheppub}

\usepackage{bbding}
\usepackage{pifont}
\usepackage{wasysym}
\usepackage{wrapfig}
\usepackage{amsmath}
\usepackage{verbatim}
\usepackage{amssymb}
\usepackage{inputenc}
\usepackage{textcomp}
\usepackage{appendix}
\usepackage{mathrsfs}
\usepackage{float}
\restylefloat{table}

\newcommand{\e}{\epsilon}

\newcommand{\be}[1]{\begin{equation}\label{#1} }
\newcommand{\ee}{\end{equation}}
\newcommand{\bea}[1]{\begin{eqnarray}\label{#1} }
\newcommand{\eea}{\end{eqnarray}}
\newcommand{\p}{\partial}
\newcommand{\refb}[1]{(\ref{#1})}
\newcommand{\N}{{\mathcal{N}}}

\newcommand{\J}{\mathcal{J}}
\newcommand{\bJ}{\bar{\mathcal{J}}}
\newcommand{\K}{\mathcal{K}}
\renewcommand{\L}{\mathcal{L}}
\newcommand{\bL}{\bar{\mathcal{L}}}

\newcommand{\z}{{\bar z}}

\renewcommand{\>}{\rangle}
\newcommand{\<}{\langle}
\newcommand{\w}{\omega}
\newcommand{\bw}{\bar{\omega}}
\newcommand{\eps}{\varepsilon}

\newcommand{\mn}{{\mu\nu}}

\newcommand{\za}{|0\rangle_\a}
\newcommand{\zc}{|0\rangle_c}

\newcommand{\spu}{|\phi\rangle}

\renewcommand{\a}{\alpha}
\newcommand{\ta}{\tilde{\alpha}}

\newcommand{\C}{\tilde{C}}
\renewcommand{\b}{\beta}
\renewcommand{\t}{\tau}
\newcommand{\s}{\sigma}
\newcommand{\cc}{\tilde{\mathscr{C}}}
\newcommand{\bes}{\begin{subequations}}
\newcommand{\ees}{\end{subequations}}
\newcommand{\ML}{\mathcal{L}}

\title{A Tale of Three\\ \vspace{0.3cm}  \Large{Tensionless strings and vacuum structure.}}

\author[a]{Arjun Bagchi,} \author[b]{Aritra Banerjee,} \author[c]{Shankhadeep Chakrabortty,}\author[a]{Sudipta Dutta,}\author[a]{and Pulastya Parekh.} \author[]{\\}

\affiliation[a]{Indian Institute of Technology Kanpur, Kanpur 208016, INDIA.\\} 

\affiliation[b]{Asia Pacific Center for Theoretical Physics, Postech, Pohang 37673, KOREA.\\}

\affiliation[c]{Indian Institute of Technology Ropar, Rupnagar, Punjab 140001, INDIA.\\}

\emailAdd{abagchi@iitk.ac.in, aritra.banerjee@apctp.org, s.chakrabortty@iitrpr.ac.in, dsudipta@iitk.ac.in, pulastya@iitk.ac.in}

\abstract{Within the premise of canonical quantisation, we re-examine the quantum structure of bosonic tensionless string theory. In the classical theory, the worldsheet metric degenerates and the Bondi-Metnzer-Sachs (BMS) algebra arises as the residual symmetries on fixing the tensionless equivalent of the conformal gauge. In the quantum regime, we find, on careful examination, that there are multiple ways to impose constraints to restrict the physical Hilbert space, which in turn lead to three distinct choices of tensionless vacua. 
We analyse these vacua in detail, commenting on various aspects like the central charges and the spectrum around each of them.}

\begin{document}
\maketitle
\section{Introduction}
All fundamental theories of Nature are inherently quantum mechanical. Classical physics, which rules most of our day-to-day experiences, emerges out of 
a classical or $\hbar \to 0$ limit of quantum mechanics. It is thus conceivable that several different quantum mechanical systems have the same classical structure, or put differently, depending on how one quantises a classical system, the resulting quantum theories can be very different. We will see this phenomenon played out in the context of the theory of tensionless strings in this paper. 
{\em A Tale of Three} is thus the story of the emergence of three different quantum tensionless string theories from a single classical closed bosonic tensionless theory.

\medskip

\subsection*{The classical picture of tensionless strings}  
String theory has been, by far, the most successful framework to understand the quantum theory of gravity. The fundamental object in string theory is one dimensional spatially extended relativistic string characterized by its tension ($T$)
\be{} T = \frac{1}{2 \pi \alpha'}.
\ee 
Here $\alpha'$ is  square of the length of fundamental string. Since the tension of fundamental string is the only \textit{free} parameter in non-interacting string theory, it is of interest to explore the two diametrically opposite limits of the theory governed by the string tension. The $T\to\infty$ limit has been widely explored and this is where the fundamental string reduces to a point particle and we recover known physics. This limit gives us low energy supergravity and a door to physics governed by ordinary quantum field theory. The opposite limit, viz. $T \to 0$, has been the source of much intrigue ever since its first exploration by Schild \cite{Schild:1976vq}. This tensionless limit is expected to probe the very highly energetic sector of the theory, where as opposed to the point-particle limit, the stringy nature of string theory becomes manifest. The hope is that this may provide a better understanding of the fundamental structure of the string theory.
 
Our study of tensionless string is based on two different approaches, which we call the intrinsic approach, following \cite{Isberg:1993av}, and the limiting approach \cite{Bagchi:2013bga, Bagchi:2015nca}. The intrinsic approach perceives the tensionless string as a fundamental object. The metric on the worldsheet of tensionless string turns out to be degenerate and hence the tensionless string is also the string equivalent of a massless point particle, i.e. a null string. The classical analysis of tensionless string is governed by a worldsheet action that naturally incorporates this degenerate metric structure and remains invariant under worldsheet  diffeomorphism \cite{Isberg:1993av}. The diffeomorphism invariance is a gauge symmetry on the worldsheet and it can only be partially fixed in a way that is analogous to choosing the conformal gauge in the tensile theory. Again, similar to the tensile theory, the gauge-fixed tensionless action still remains invariant under a residual gauge symmetry structure on the worldsheet. Interestingly, instead of two copies of the Virasoro algebra in the tensile case, now the generators of this residual gauge symmetry close to form the Bondi-Metzner-Sachs (BMS) algebra, which was originally proposed as the asymptotic symmetry algebra of Minkowski spacetime \cite{Bondi:1962px, Sachs:1962wk}. Starting from the BMS invariant gauge fixed action of tensionless strings, one can systematically work out the equation of motion, mode expansions and constraint analysis of the theory \cite{Isberg:1993av, Bagchi:2015nca}.

On the other hand, in the limiting approach of tensionless analysis, we begin with the usual tensile string theory and argue that the tensionless limit is one where the fundamental string becomes very long and floppy. In terms of worldsheet coordinates $\{\t, \s\}$ of a closed bosonic string, this is the scaling $\{\tau \to \epsilon \tau, \sigma \to \sigma\}$ with the scaling parameter $\epsilon \to 0$ \cite{Bagchi:2013bga}. This can also be viewed as an infinite boost which makes the worldsheet null. The limit described above sets the velocity of light on the worldsheet to be zero \cite{Bagchi:2012cy}. The results obtained in the intrinsic analysis of tensionless string can be recovered by taking this ultra-relativistic limit on worldsheet of closed bosonic tensile strings \cite{Bagchi:2015nca}. Specifically, scaling the residual symmetry algebra in the closed bosonic tensile theory, viz. the 2d conformal algebra, we get the $\text{BMS}_3$ algebra. This, as we just mentioned above, is the residual symmetry algebra as obtained in the intrinsic analysis of tensionless string theory. Such comparison between intrinsic analysis and limiting approach can also be extended to supersymmetric theory. It has been shown, two distinct ultra-relativistic scalings on $\mathcal{N} = (1,1)$, 2d super conformal algebra as the residual symmetry algebra of the tensile closed superstring theory, give rise to both homogeneous and inhomogeneous versions of super $\text{BMS}_3$ algebra \cite{Bagchi:2016yyf, Bagchi:2017cte}. 

Interestingly, instead of taking the ultra-relativistic limit, if we consider a non-relativistic limit on the residual symmetry algebra in tensile closed bosonic string theory by defining a scaling $\tau \to  \tau, \sigma \to \epsilon \sigma$ and then sending the scaling parameter $\epsilon \to 0$, we again get the $\text{BMS}_3$ algebra \cite{Bagchi:2013bga}. However, it is yet to be fully understood how to invent an exhaustive intrinsic analysis that begins with a fundamental action and perfectly fits with the limiting picture appearing from the non-relativistic limit in the usual tensile case.

\subsection*{Quantum tensionless strings} 
We now understand the classical theory of the tensionless strings reasonably well. We will review the construction in Sec. \ref{four}. The question we attempt to answer in this paper is how we should go about quantising this theory. Before focusing on the details of this paper, let us briefly mention why this may be of interest. 

It has been investigated long back that in a very high energetic regime of the perturbative quantum tensile string theory, the scattering amplitudes starts behaving in way so that together they satisfy infinite number of linear relations among themselves and that indicates the emergence of a symmetry structure which is expected to be much larger than the existing symmetries of the usual tensile string theory. An appropriate realisation of such a high energetic sector of string theory was speculated to be the tensionless limit in the theory \cite{Gross:1987kza, Gross:1987ar, Gross:1988ue}. 

The strange behaviour of perturbative degrees of freedom of string theory at extreme conditions is manifested in another occasion as we approach a very high temperature known as Hagedorn temperature $\mathcal T_H$ \cite{Bowick:1989us, Giddings:1989xe}, where the string partition function diverges. Near this extreme temperature,  string theory is best described by its tensionless limit as the effective tension of the fundamental string gradually approaches zero \cite{Pisarski:1982cn, Olesen:1985ej}. It has been speculated \cite{Atick:1988si} that beyond the Hagedorn temperature, string theory undergoes a phase transition and the fundamental degrees of freedom in the new phase of the theory is very different from the usual perturbative degrees of freedom.  

Also, unlike the case of usual perturbative description of tensile theory, a precise bound on the critical dimension of string theory at the zero tension limit is yet to be fully understood {\footnote{There are a variety of answers in the literature, many of which apparently contradict each other, for example see \cite{Karlhede:1986wb, Lizzi:1986nv, Gamboa:1989px}.}}. Therefore a better understanding of quantum theory of free tensionless strings from first principles is essential to address all these intriguing issues of string theory at this extreme limit.

\subsection*{Our present work} 
In this paper, we address the problem of quantisation of the classical closed bosonic tensionless string theory. We will confine ourselves to the domain of canonical quantisation and will discover that there are multiple quantum mechanical theories that emerge from the same classical tensionless theory. 

The analysis of constraints is the principle factor in determining the quantum structure of the tensionless string theory. In the present work, we start with classical constraints arising from the non vanishing component of energy momentum (EM) tensor present in the gauge-fixed version of the intrinsic, tensionless string theory. For all of our subsequent analyses, we use the same set of classical constraints expressed in terms of the Fourier modes of the EM tensor. Now, once the theory is quantized, we need to impose the quantum version of those constraints on each state that belongs of the tensionless Hilbert space. The imposition of quantum constraints is realised as the vanishing of the matrix elements of the Fourier mode of EM tensor, with respect to the all possible states in tensionless Hilbert space. We will call this way of imposing quantum constraints the $sandwich$ conditions. 

In this work, we look for all possible consistent ways to implement such the quantum $sandwich$ conditions on the intrinsic tensionless theory. Starting from a unique classical action as well as the underlying classical symmetry algebra, we show that there are nine inequivalent ways to implement the $sandwich$ conditions in the intrinsic analysis of quantum tensionless strings. However, by analyzing the compatibility between the implementation of those conditions and the underlying quantised version of the symmetry algebra we can rule out some such possibilities. Ultimately, we are left with {\em three} inequivalent cases and interestingly they correspond to three distinct tensionless vacua and hence three very different quantum theories. One of the vacua is the so-called induced vacuum, the vacuum of the induced representation of the underlying BMS algebra. This also follows from the ultra-relativistic limit of the string worldsheet and has intriguing features like the emergence of a long open string, as was recently described in \cite{Bagchi:2019cay}. Another vacuum, which we call the flipped vacuum, leads to the bosonic version of the Ambitwistor string theory. Connections between ambitwistor strings and the tensionless limit have been discussed recently in \cite{Casali:2016atr}. The third vacuum, which we name the Oscillator vacuum, is completely new and contains intriguing hints of a huge underlying gauge symmetry, massless higher spins, as well as a massive sector. We go on to describe these three vacua and their properties in some detail in the rest of this paper. 

The organization of this paper is as follows: In section \ref{two}, we remind the reader about BMS$_3$ algebra, how it appears as different contractions of the 2D Virasoro algebra and related representation theory. In section \ref{three}, we review some classical and quantum aspects of tensile closed bosonic string theory which we consider very relevant for all subsequent sections of the paper. In section \ref{four}, we present an extensive discussion on the classical tensionless string, both from the point of views of intrinsic picture and the limiting side.  Section \ref{five} plays the pivotal role in this paper as because it explains how we impose the quantum constraints on the states in quantum tensionless string and also how the implementation of such quantum constraints leads to different inequivalent descriptions of quantum tensionless string. The next three sections are dedicated to the discussions of the three inequivalent vacua in some detail. We end with some conclusions and future directions. There are a couple of appendices the first of which deals with the inconsistent cases of imposition of quantum constraints. The other one details the computation of central charges in the various vacua. 

\section{Algebraic Preliminaries}
\label{two}
\subsection{The Virasoro algebra} 


It is well known that in two dimensions the conformal algebra is infinite dimensional and consists of two copies of the Virasoro algebra. On a complex plane mapped by $z$ and $\bar{z}$ coordinates, the generators of the infinite set of transformations are written as
\be{}
\ML_n=z^{n+1}\p_z;\quad \bL_n=\bar{z}^{n+1}\p_{\bar{z}}. \\
\ee{}
They generate two copies of the Virasoro algebra:
\bea{VIRC} [\ML_m,\ML_n]&=&(m-n)\ML_{m+n}+\frac{c}{12}m(m^2-1)\delta_{m+n,0}, \cr
~~ [\ML_m,\bL_n]&=&0,\cr
~~ [\bL_m,\bL_n]&=&(m-n)\bL_{m+n}+\frac{\bar{c}}{12}m(m^2-1)\delta_{m+n,0}. 
\eea
The classical algebra is given without the central charges $c$ and $\bar{c}$. The central terms arise due to quantum anomalies and ordering ambiguities. The form $m(m^2-1)\delta_{m+n,0}$ is obtained by imposing Jacobi identities.
\subsection*{Highest weight representations of the Virasoro algebra}

The two dimensional Virasoro algebra can be realised on a state space that is spanned by the eigenstates of $\ML_0$ and $\bL_0$ operators, with eigenvalues $h$ and $\bar{h}$:
\bea{two.8}
\ML_0|h,\bar{h}\>&=&h|h,\bar{h}\>;\quad \bL_0|h,\bar{h}\>=\bar{h}|h,\bar{h}\>,
\eea
The commutation relations \refb{VIRC} gives us 
\be{two.9} \ML_0\ML_n|h,\bar{h}\>=(h-n)\ML_n|h,\bar{h}\>;\quad \bL_0\bL_n|h,\bar{h}\>=(\bar{h}-n)\bL_n|h,\bar{h}\>. \ee
This tells us that the $\ML_n$ and $\bL_n$ reduces the corresponding weight by $n$. Now, if we demand that there exists a state $|h,\bar{h}\>$ which is annihilated by the positive modes of the generators:
\be{11.6} \ML_n |h,\bar{h}\>=\bL_n|h,\bar{h}\>=0\quad (n>0), \ee
we obtain a spectrum that is bounded from below. Therefore we define $|h,\bar{h}\>$ as the primary state, and from this state the descendants are given by the action of the negative modes $ \ML_{-n}$ or $ \bL_{-n}$ (for $n>0$). The primary state together with its descendants form the highest weight representation of the algebra. The residual worldsheet symmetries of  bosonic tensile closed strings generate two copies of the Virasoro algebra, and the physical states are usually classified using the highest weight representations.


\subsection*{An automorphism}\label{flippy}

Curiously, there could be other representations of the Virasoro algebra. The two dimensional algebra can be seen to admit an automorphism \cite{Bagchi:2019unf} given by 
\be{automorph}
\bar{\mathcal{L}}_n\rightarrow\bar{\mathcal{L}}'_n=-\bar{\mathcal{L}}_{-n}.
\ee
If we start from the Virasoro generators and perform the above transformation in the anti-holomorphic sector, keeping the holomorphic ones preserved, the algebra remains preserved. The existence of this automorphism means there is a `flip' between the raising and lowering operators in one sector of the theory. 
If one also considers the centrally extended version, the algebra will again remain invariant under above transformations provided we have, 
\be{}
 \bar{c} \to \bar{c}' = - \bar{c}.
\ee
Under this automorphism, the highest weight representation in the anti-holomorphic sector becomes a lowest weight representation. This is given by $\bar{h}\rightarrow\bar{h}'=-\bar{h}$. One can still define states, which are given by 
\begin{subequations}\label{flippy1}
\bea{}
\ML_n|h,\bar{h}\>&=&\bar{\ML}_{-n}|h,\bar{h}\>=0\ \forall\ n>0, \\
(\ML_0-h)|h,\bar{h}\>&=&(-\bL_0+\bar{h})|h,\bar{h}\>=0.
\eea
\end{subequations} 
We can call this as the ``flipped'' representation. The fact that there exists a string theory where worldsheet symmetries correspond to Virasoro algebra, but the physical states are defined in the above way, makes this representation interesting. We will come back to this theory in the next section.

\subsection{Taking limits on Virasoro} 

Non-relativistic algebras arise from contraction of relativistic ones at the level of generators.
An In{\"{o}}n{\"{u}}-Wigner contraction of the $D$ dimensional relativistic conformal algebra results in a finite dimensional Galilean Conformal Algebra (GCA) \cite{Bagchi:2009my}. The contraction of the conformal algebra \refb{VIRC} can be achieved in two ways: either by considering a non-relativistic (NR) limit, or an ultra-relativistic (UR) limit on the generators. In two dimensions, the 2D GCA is known to be isomorphic to the Bondi-Metzner-Sachs (BMS) algebra in 3D. This is also known as the BMS$_3$/GCA$_2$ correspondence \cite{Bagchi:2010eg}. Given that the 2D GCA is a contraction of the Virasoro algebra, this means that the symmetry structure of flat space could be understood as a limit of the symmetry structure of $AdS_3$ space \cite{Bagchi:2012cy, Barnich:2012aw}. This naturally provided support for the idea that flat holography could indeed be understood as a limit of usual AdS$_3$/CFT$_2$. 


Unlike in the case of the conformal algebra, the GCA can be given an infinite dimensional lift even for $D\geq 2$. This striking feature of GCA has been  extensively studied in non-relativistic field theories such as Galilean electrodynamics \cite{Bagchi:2014ysa} and later in non-relativistic versions of Yang-Mills theories \cite{Bagchi:2015qcw,Bagchi:2017yvj}. We get the same number of generators that obey the GCA.  As it happens to be, in the case of string worldsheet theory, the two possible contractions are physically very intriguing. In the NR case, the worldsheet speed of light is scaled to infinity: $c\to\infty$, while in the UR case, the speed of light is scaled to zero. Since we are interested in the case of 2D, both these contractions generate the same algebra at the classical level \cite{Duval:2014uoa}. 

In the NR limit, the generators of the new algebra are related to the conformal ones by the following scaling
\bea{cont1}
L_n=\ML_n+\bL_n; ~~
M_n=\e(\ML_n-\bL_n), ~~\e\to0
\eea
The commutators between the generators can be worked out and it can be found that this generates the algebra,
\bea{BMS3}
&& [L_n, L_m] = (n-m) L_{n+m} + \frac{c_L}{12} \, (n^3 -n) \delta_{n+m,0}, \nonumber\\
&& [M_n, M_m] = 0, \nonumber \\
&& [L_n, M_m] = (n-m) M_{n+m} + \frac{c_M}{12} \, (n^3 -n) \delta_{n+m,0}. 
\eea

 In the above $L_n$ are the generators of diffeomorphisms of the circle at the null infinity (super-rotations) and $M_n$ are the generators of super-translations. Classically the algebra is without the central extension. However, if one were to have central charges, they would be related to the Virasoro central charges by 
\be{} c_L = c + \bar{c};\quad c_M= \e ( c - \bar{c} ).\ee
When the original CFT has $c = \bar{c}$, then the central charge goes to $c_M=0$. On the other hand, the UR contraction is achieved by performing the following scaling on the cylinder:
\bea{cont2}
L_n=\ML_n-\bL_{-n};~~
M_n=\e(\ML_n+\bL_{-n}), ~~\e\to0
\eea
Notice here the mixing between positive and negative modes during the contraction, which is particular to the UR case. It can be checked that this too generates the BMS$_3$ algebra \refb{BMS3}. In the UR limit the central charges are related by 
\be{two.26} c_L = c - \bar{c};\quad c_M= \e ( c + \bar{c} ).\ee
 Thus, if we are interested in the limit of a field theory having $c=\bar{c}$, it must have vanishing $c_L$.  We also note that to keep $c_M$ finite, we would need to scale $c, \bar{c}$ to be infinitely large.

\subsection{Representation theory of BMS Algbera} \label{BMSrep}
 Intrinsically speaking, the Bondi-Metzner-Sachs (BMS) group is a vital toolkit in understanding holography in asymptotically flat spacetimes. Borrowing wisdom from conventional AdS/CFT, the 2D field theory dual to 3D flat space should live on the null boundary and inherit the BMS symmetries. This line of inquiry has led to recent successes \cite{Bagchi:2010eg,Bagchi:2012xr,Bagchi:2012cy,Bagchi:2012yk,Bagchi:2013lma,Bagchi:2014iea,Riegler:2014bia,Grumiller:2019xna} in flat holography. The fact that BMS$_3$ symmetry arises as the worldsheet symmetry of tensionless strings, is indeed remarkable in this sense. One could take advantage of the known machinery of BMS representations to formulate such tensionless worldsheet theories. 

\subsection*{Highest weight representations }

Let us take a look at the highest weight representations of the BMS$_3$ Algebra \refb{BMS3}. Keeping in with the Virasoro case, we will label the states with the eigenvalues of $L_0$, which we will call $h_L$. Since $[L_0, M_0]=0$, this means that these states can be further labeled by $h_M$, the eigenvalue of $M_0$. Therefore 
\bes\bea{} L_0|h_L,h_M\>&=&h_L|h_L,h_M\> \\
M_0|h_L,h_M\>&=&h_M|h_L,h_M\>. \eea\ees
Using the commutators of the BMS algebra, we get the following relations
\bes\bea{} L_0 L_n|h_L,h_M\>&=&(h_L-n)L_n|h_L,h_M\>\\
 L_0 M_n|h_L,h_M\>&=&(h_L-n)M_n|h_L,h_M\>. \eea\ees
We observe that $L_{-n}$ or $M_{-n}$ raises the weight $h_L$ by $n$. We can give the notion of a primary state to $|h_L, h_M\>$ in a theory where the value of $h_L$ is bounded from below. This is achieved by demanding the following conditions 
\be{11.14} L_n|h_L,h_M\>=M_n|h_L,h_M\>=0\quad (n>0). \ee
Starting from the primary state $|h_L,h_M\>$ we can build up a tower of states by the action of $L_{-n}$ and $M_{-n}$ with $n > 0$. These are the  descendants of the primary, therefore giving a representation of BMS$_3$. 

\subsection*{Induced representations } \label{2.2.2}

Let us consider a particular state $|M,s\>$ in the Hilbert space which satisfies:
\begin{subequations}\label{ind}
\bea{}
&& M_0 |M,s\> = M |M,s\>, \, L_0 |M,s\> = s |M,s\>; \\ 
&& M_n |M,s\> = 0, \ \forall n \neq 0.
\eea
\end{subequations}
This defines a one dimensional representation of the sub-algebra of BMS$_3$ spanned by $\{L_0, M_n, c_L, c_M\}$. This can be used to define an {\em{induced BMS module}} with basis vectors 
\be{}
|\Psi\>= L_{n_1} L_{n_2} \ldots L_{n_k} |M,s\>.
\ee 
Here $n_1\geq n_2 \geq \ldots \geq n_k$ are integers which can be both positive or negative. These induced BMS representations have been studied recently \cite{Barnich:2014kra, Barnich:2015uva, Campoleoni:2016vsh}. An important class of representations for the BMS$_3$ algebra is the so-called {\em{massive modules}} \cite{Campoleoni:2016vsh}. 

\subsubsection*{Mapping of representations}

Lastly, we have to take a closer look at the mapping between the Virasoro and BMS generators under the NR and UR limits as given in  equations \refb{cont1} and \refb{cont2}. We observed earlier that unlike the NR limit, there is a mixing of positive and negative modes of Virasoro generators in the UR limit. This becomes an important fact while mapping the representations. The NR limit on \refb{11.6} results in 
\be{nrrep}
\left(L_n + \frac{1}{\e} M_n \right)|h, \bar{h} \> = \left(L_{n} - \frac{1}{\e} M_{n} \right) |h, \bar{h} \> =0, \ n>0,
\ee
which can be reduced to \refb{11.14} by identifying $|h, \bar{h} \>=|h_L, h_M \>$. Therefore, we observe that the highest weight states of Virasoro map to highest weight states of BMS in the NR limit. 
 However, in case of the UR limit, the Virasoro highest weight representations map to induced representations of BMS. 
The UR limit on \refb{11.6} can be seen to translate into 
\be{}
\left(L_n + \frac{1}{\e} M_n \right)|h, \bar{h} \> = \left(L_{-n} - \frac{1}{\e} M_{-n} \right) |h, \bar{h} \> =0, \ n>0. 
\ee
If we assume that the Virasoro highest weight state $|h, \bar{h} \>$ in the limit $\e\to0$ maps the following way
\be{ms} \lim_{\e\to0} |h, \bar{h} \> = |M,s\>,\ee
then the conditions \refb{two.8}, \refb{two.9} automatically give rise to
\begin{subequations}\label{ind}
\bea{}
 M_0 |M,s\> = M |M,s\>, \, ~L_0 |M,s\> = s |M,s\>; ~~~
M_n |M,s\> = 0, \ \forall\  n \neq 0,
\eea
\end{subequations}
along with the identification $M = \frac{\e}{2} (h + \bar{h})$ and $s= \frac{1}{2}(h - \bar{h})$. These conditions are nothing but that of induced representations as described earlier. When we study the induced vacuum of the tensionless string in Sec~\ref{Iv} we will see how these induced representations play a big role in its formulation. 

The other intriguing take home message here is the fate of flipped representations, as discussed in  (\ref{flippy1}), under such contractions. The reader may already notice that the NR and UR contractions are related to each other via the automorphism 
(\ref{automorph}). One can show that the UR limit of such a flipped representation effectively acts as the NR limit on the theory. In the recent work \cite{Bagchi:2018wsn} we elucidated more on this issue, and explored its consequences. It comes as no surprise that a limit from a string theory based on the flipped representations would give rise to BMS$_3$ worldsheet symmetries and primaries of Virasoro would naturally  map to primaries of BMS$_3$ as in (\ref{nrrep}).

\begin{figure}[H]
\begin{center}
{\includegraphics[scale=0.4]{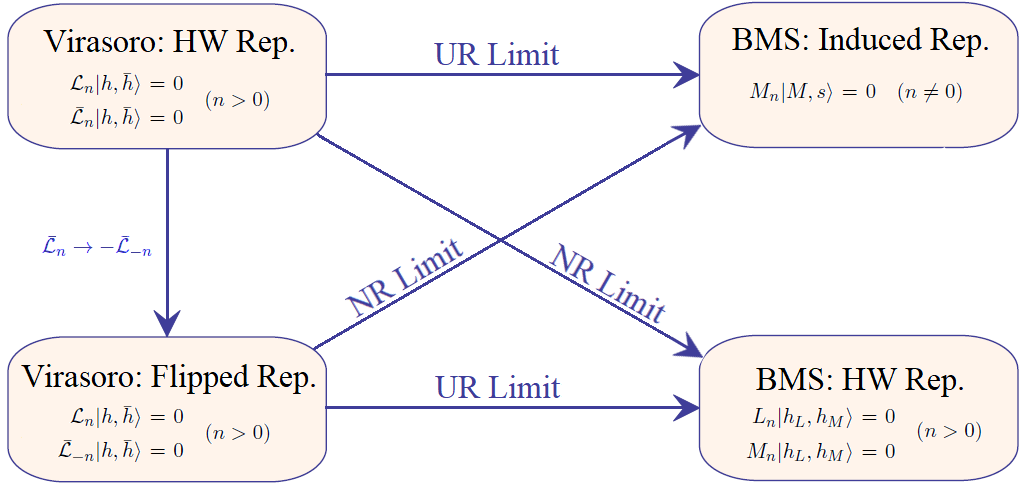}}
\caption{Mapping of different representations in the singular limits.}\label{fig44}
\end{center}
\end{figure}

All in all, there is a web of relations between the representation theories of 2D conformal algebras and the GCA$_2$/BMS$_3$ that arises in the limit. Such a web may be summarised by Figure \ref{fig44}, which will be crucial to our discussion throughout this work.


\section{Back to school: quantum bosonic tensile string}
\label{three}

\subsection{The ``usual'' string}

In this section we are going to quickly review the classical and quantum aspects of closed bosonic tensile string theory. The classical dynamics of tensile bosonic string is governed by the Polyakov action,
\be{weyl}
S = -\frac{T}{2}  \int d^2 \xi  \sqrt{-g} g^{\alpha \beta} \p_\alpha X^{\mu}(\tau,\sigma) \p_\beta X^{\nu}(\tau,\sigma) \eta_{\mu \nu},
\ee
where $T$ is the string tension, $g_{ab}$ is the intrinsic metric on the worldsheet spanned by the coordinates $\xi^{\a}=(\t,\s)$ and $X^{\mu}(\tau,\sigma)$ is the embedding of the worldsheet in the background spacetime $\mathbb{R}^{1,d-1}$. The Polyakov action is preserved under symmetries on the world sheet, e.g, the reparameterization invariance $ \xi^{\alpha} \to \xi'^{\alpha}(\xi)$,  and the Weyl invariance $g^{\alpha \beta} \to e^{\phi(\tau,\sigma)}g^{\alpha \beta} $. The equations of motion with respect to $X^{\mu}(\tau,\sigma)$ and $g^{\alpha \beta}$, derived from the action (\ref{weyl}), are as follows,
\be{eom}
\partial_{\alpha} (\sqrt{-g} g^{\alpha \beta} \partial_{\beta}X^{\mu}) = 0,~~~~~~~ T_{\alpha \beta} = \partial_{\alpha}X^{\mu} \partial_{\beta}X_{\mu} - \frac{1}{2} g_{\alpha \beta} g^{\alpha' \beta'} \partial_{\alpha'}X^{\mu} \partial_{\beta'}X_{\mu} = 0,
\ee
where $T_{\alpha \beta}$ is the worldsheet energy momentum tensor. By exploiting the freedom of reparameterization invariance and the Weyl invariance one can suitably choose conformal gauge, by setting  $g^{\alpha \beta} = \eta^{\alpha \beta}$.  As a consequence, $T_{\alpha \beta} = 0$ becomes purely classical constraint equation. In this gauge, the form of equations of motion and constraints in terms of worldsheet lightcone coordinates ($\s^\pm=\t\pm\s$) simplifies as,
\be{}\p_+\p_-X^\mu=0;\quad\quad T_{++} = (\p_+X)^2= 0;\quad\quad T_{--} =(\p_-X)^2=0. \ee
The mode expansion compatible with the periodic boundary condition of closed bosonic string theory, $X^{\mu} (\tau, \sigma+ 2 \pi) = X^{\mu} (\tau, \sigma)$, turns out to be,
\be{tensilemode}
  X^\mu(\t,\s)=x^\mu+2\sqrt{\frac{\a'}{2}}\a^\mu_0\t+i\sqrt{\frac{\a'}{2}}\sum_n\frac{1}{n}\Big[\a^\mu_n e^{-in(\t+\s)}+\tilde{\a}^\mu_n e^{-in(\t-\s)}\Big],
   \ee
 where $\a^\mu_n (\tilde{\a}^\mu_n)$ are the left (right) moving classical oscillators, and $x^{\mu}$ is the position of centre of mass for closed bosonic string.  The reality condition on $ X^\mu(\t,\s)$ is manifested as $(\a^\mu_n)^\dagger=\a^\mu_{-n}$ and $(\ta^\mu_n)^\dagger=\ta^\mu_{-n}$. The zero modes are related to the momentum ($k^\mu$) by $\a^\mu_0=\ta_0^\mu=\sqrt{\frac{\a'}{2}}k^\mu$. Re-expressing the classical constraints in terms of the Fourier modes of the non vanishing components of energy momentum tensor, 
\bea{} T_{++}=2\a'\sum_n \ML_n\ e^{-in\s^+} ,~~~ T_{--}=2\a'\sum_n \bL_n\ e^{-in\s^-}, \eea
the constraint equation takes  simplified form, 
\be{} \ML_n = 0, ~~ \bL_n = 0 \quad\forall\ n. \ee  
$\ML_n$,$\bL_n$ are given in terms of oscillators by 
\be{} \ML_n=\frac{1}{2}\sum_m \a_{-m}\cdot\a_{m+n},~~
\bL_n=\frac{1}{2}\sum_m \ta_{-m}\cdot\ta_{m+n} \quad\forall\ n, \ee
where the sum is over all integers. In conformal gauge, the worldsheet theory can be interpreted as a conformal field theory of $D$ independent scalars, $X^{\mu}(\xi)$ in two dimensional flat spacetime. Following the standard field theoretic technics, one can quantize the worldsheet theory  by defining a $tensile~vacuum$. The canonical commutation relations among the oscillator modes are given as:
\be{} [\a^\mu_m,\a^\nu_n]=m\delta_{m+n}\eta^\mn, 
~~ [\ta^\mu_m,\ta^\nu_n]=m\delta_{m+n}\eta^\mn, 
~~ [\a^\mu_m,\ta^\nu_n]=0.
 \ee
The Hilbert space is built by acting the creation operators $\a^\mu_{-n}$, $\ta^\mu_{-n}$ $(n>0)$ on the tensile vacuum $|0, k^{\mu}\rangle_\a$ defined in the following way,
\be{alphvac}
\a^\mu_n|0, k^{\mu}\rangle_\a=\ta^\mu_n|0, k^{\mu}\rangle_\a=0,\quad\forall\ n>0. 
\ee
A tower of excited states can be built up over the vacuum and all of them including the tensile vacuum have  well-defined energy(mass). Finally one needs to impose a stronger version of quantum constraint as the physical state condition on the elements of the Hilbert space of the theory. 
\be{quantconstr}
\langle phys| T_{\alpha \beta} | phys'\rangle = 0 
\ee
For closed bosonic string, the above quantum constraint (\ref{quantconstr}) further modifies to, 
\be{fmodecond} \<phys'|\ML_n|phys\>=\<phys'|\bL_n|phys\>=0,\quad\forall\ n  \ee

The implementation of the stronger version  of the quantum constraint (\ref{fmodecond}) gives rise to the right hand action of operators $ \ML_n, \bL_n$ on the physical states. 
\bea{implement}
\ML_n|phys\>&=&\bL_n|phys\>=0,\quad\forall\ n>0, \nonumber\\
(\ML_0-a)|phys\>&=&0,~~(\bL_0-\bar{a})|phys\>=0, \eea

 where $a$ and $\bar{a}$ are the normal ordering constants. Due to the non trivial commutators between the creation and annihilation oscillators, we need to normal order $\ML_0$ and $\bL_0$ in the above equations, where would have $a$ and $\bar{a}$ as normal ordering ambiguities. Explicitly, the normal ordered operators are
\bes \label{zeromode}\bea{} \ML_0&=&\frac{1}{2}\a_0^2+\sum_{m>0}\a_{-m}\cdot\a_{m}=\frac{1}{4}\a'k^2+N_L, \\
\bL_0&=&\frac{1}{2}\ta_0^2+\sum_{m>0}\ta_{-m}\cdot\a_{m}=\frac{1}{4}\a'k^2+N_R. \eea \ees
$N_L$ and $N_R$ counts the number of left and right modes in the states. The zeroth mode is related to $k^\mu$, the momentum of the string, and $\a'=\frac{1}{2\pi T}$. 
$ \ML_n$ and $\bL_n$ close under two independent copies of centrally extended Virasoro algebra \refb{VIRC}.

The implementation of constraints in (\ref{implement}) naturally furnishes the highest weight representation of Virasoro algebra. The form of this algebra and the values of the charges can be calculated by applying the Jacobi identities worked out in Appendix \ref{ApA}. For the vacuum $\za$ one finds the central charges to be
\be{} c=\bar{c}=D. \ee
 The closed string Hilbert space satisfying (\ref{implement}) still suffers from having negative norm states in the spectrum. It turns out that only for particular values of these parameters $a$, $\bar{a}$ and $D$, we can remove the negative norm states from the Hilbert space. Considering vacuum $|0,k^\mu\rangle_\a$ as a physical state one can show $a = \bar{a}$. Further, by considering the first excited state as a physical state, one can show it always comes with non-negative norm only when $a$ follows an inequality $a\leq 1$ . From the physical state analysis with higher excited states we will not get any new condition on $a$.
 
To inspect the bound on $D$ for removing negative norms we need to consider \textit{spurious states} $\spu$ in the spectrum. These are special states orthogonal to the physical states: $\langle phys\spu=0$. They are attributed with the properties:
\be{11} (\ML_0-a)\spu=0,~~ (\bL_0-a)\spu=0. \ee
Since these are orthogonal to physical states, they can always be written as:
\be{} \spu\sim\sum_{n>0}\ML_{-n}|\chi_n\rangle\ or\ \sum_{n>0}\bL_{-n}|\chi_n\rangle\ee
The conditions \refb{11} for spurious states translates to 
\be{} \ML_0|\chi_n\rangle=\bL_0|\chi_n\rangle=(a-n)|\chi_n\rangle \ee
Any spurious state can be broken down in terms of $\ML_{-1},\bL_{-1}$ or $\ML_{-2},\bL_{-2}$ since $\ML_{-3}\sim[\ML_{-1},\ML_{-2}]$ etc.
At level one, we can have a spurious state of the form:
\be{} |\phi_1\rangle=\ML_{-1}\bL_{-1}|\chi_1\rangle. \ee
A most general level two spurious state will be of the form:
\be{} |\phi_2\rangle=(\ML_{-2}\bL_{-2}+b_1\ML_{-1}\ML_{-1}\bL_{-2}+b_2\ML_{-2}\bL_{-1}\bL_{-1}+b_3\ML_{-1}\ML_{-1}\bL_{-1}\bL_{-1})|\chi_2\rangle, \ee
where the $b_i$'s are constants. If we demand that the states $|\phi_1\rangle$ and $|\phi_2\rangle$ are $spurious$ and $physical$ simultaneously, then we will have such states with zero norm, since they are perpendicular to themselves. That amounts to imposing the following conditions and equations on $|\phi_1\rangle$ and $|\phi_2\rangle$:
\be{} \ML_0|\chi_1\rangle=\bL_0|\chi_1\rangle=(a-1)|\chi_1\rangle,~~
 \ML_0|\chi_2\rangle=\bL_0|\chi_2\rangle=(a-2)|\chi_2\rangle, \nonumber \ee
\be{} \ML_1|\phi_1\rangle=\bL_1|\phi_1\rangle=0,~~
 \ML_2|\phi_1\rangle=\bL_2|\phi_1\rangle=0. \ee
If we use the Virasoro algebra with central term along with above conditions we find that  $a=1$, $b_1=b_2=\frac{3}{2}$, $b_3=\frac{9}{4}$ and $D=26$ form a set of conditions to ensure $ |\phi_2\rangle$ is a null state. We can further check that physical state with negative norm is possible for $D > 26$. However, the physical spectrum becomes free of such negative norm state for  $a =1$, $D=26$ and $a \leq 1$, $D \leq 25$. It is beyond the scope of free theory to ascertain that the ghost free theory must live in $D=26$ dimensions. These ideas and details are well known, but will be absolutely necessary to keep in mind as we go forward to tensionless vacua, where the relevant physical conditions work quite non-trivially.

\subsection{The ``flipped'' string} 
\label{F11}

Here we move on towards briefly discussing the salient features of the quantum theory that follows from the choice of a flipped vacuum as we described in connection to flipped representations of the Virasoro algebra (\ref{flippy1}). This will help us to understand the highest weight representations of the states in the tensionless limit, which we are going to consider in section \refb{flip}. Let us start by defining a twisted string vacuum by 
\be{flipvac}
\a_n|0\>_A=\ta_{-n}|0\>_A=0\quad\forall\ n>0.
\ee
Then the analysis outlined in the previous section follows. Here we are only going to mention the deviation from the usual case and their consequences. As seen from \refb{automorph}, the constraint $\bL_n$ picks up a sign in this case. Since the creation and annihilation operators are interchanged in one sector, the ordering of $\bL_0$ has to be adjusted accordingly. Therefore the zero modes of the Virasoro constraints \refb{zeromode} are now given by 
\bes\label{} \bea{} \ML_0&=&\frac{1}{4}\a'p^2+N_L, \\
\bL_0&=&-\frac{1}{4}\a'p^2+N_R. \eea \ees
where the definition of the number operator also changes: $N_R=\sum_{m>0}\ta_{m}\cdot\ta_{-m}$.  
The constraint condition is imposed on the physical states by the familiar condition on matrix elements,
\be{sand2} \<phys'|\ML_n|phys\>=\<phys'|\bL_n|phys\>=0,\quad\forall\ n. \ee
The creation and annihilation modes of $\bL$ are now flipped as well. So we realise this condition by considering the same ones as in  (\ref{flippy1}),
\bea{1pp}
\ML_n|phys\>&=&\bL_{-n}|phys\>=0,\quad\forall\ n>0, \nonumber\\
(\ML_0-a)|phys\>&=&0,~~(\bL_0-\bar{a})|phys\>=0. \eea
This is in fact a combination of the highest weight representation in the holomorphic sector and the lowest weight representation in the anti-holomorphic sector.  This is our very first encounter with unusual ways of describing physical states in a theory of string worldsheet. Adding and subtracting the equations in \refb{1pp} gives us
\bes\label{2.104}\bea{} \Big[\frac{1}{2}\a'p^2+N_L-N_R\Big]|phys\>&=&0, \\
 (N_L+N_R-2a)|phys\>&=&0. \eea\ees
For $a=1$, this tells us that instead of getting the level matching condition, we get a spectrum restricted at $N_L+N_R=2$. The mass formula on the other hand becomes  
\be{} m^2|phys\>=\frac{2}{\a'}(N_L-N_R)|phys\>. \ee
The allowed states are therefore given by the values
\be{} (N_L,N_R)=(1,1),(2,0),(0,2), \ee
with the mass squared being $0$, $\frac{4}{\a'}$ and $-\frac{4}{\a'}$ respectively. The spectrum consists of a massless tensor, and two vector bosons which are massive and tachyonic. These states turn out to be useful for computing amplitudes of massless sectors. These twisted string theories have been introduced in \cite{Huang:2016bdd} and more recently illustrated in detail in \cite{Lee:2017utr,Lee:2017crr}. 

One must note here that the tensile spectrum itself is quite subtle in this case. Particularly speaking, the massless $(1,1)$ state  $\xi_{\mu\nu}\alpha_{-1}^{\mu}\overline{\alpha}_{+1}^{\nu}|0\rangle$ gives rise to negative norm, where $\xi_{\mu\nu}$ is a polarisation tensor. It was argued in \cite{Lee:2017utr} that to make the massless sector physical, one has to consider that the vacuum has negative norm $|||0\>_A||<0$. This state is often dubbed as ``gravity'' sector of the string spectrum as it contains symmetric, anti-symmetric and trace (scalar) excitations. But in reality one can show the anti-symmetric excitation has zero norm, which is quite different from the usual string case. This abundance of extra null states in the parent theory will also cause peculiarities in the daughter tensionless theory, which we will discuss in later sections.
\subsubsection*{Null States in the flipped theory}

For completeness, we dabble with the spurious state analysis in the flipped parent theory. Since, the holomorphic sector remains same as in the usual tensile theory, we can write the left moving null state at level two as,
  \be{}
  |\chi_L\rangle = \left(\ML_{-2}+\eta \ML_{-1}\ML_{-1}\right)|\phi\rangle,
  \ee
  With $\eta =-\frac{3}{2}$.
  The antiholomorphic sector is more subtle here and we in this case demand the following to identify a null state,
  \be{}
\bL_{-1}\left(\bL_{2}+\overline\eta \bL_{1}\bL_{1}\right)|\phi\rangle=0.
  \ee
  Keeping in with the flipped nature of the theory, we can write
  $
  \overline{\eta} = \frac{3}{2},
  $
 i.e. $\eta=-\overline\eta$. So, we have the following form for the antiholomorphic null state, 
  \be{}
  |\chi_R\rangle = \left(\bL_{2}+\overline\eta \bL_{1}\bL_{1}\right)|\phi\rangle.
  \ee
These will be important when we describe null states in the tensionless version of these flipped strings.

\newpage

\section{The Classical Tensionless Story}
\label{four}
The massless limit on the point-particle action projects one to null geodesics on the background spacetime. So the metric of the worldline of the massless point particle is degenerate. In a very similar way, the tensionless limit of string theory, projects the string onto null worldsheets. This can be seen explicitly by working in the Hamiltonian framework and then writing a phase-space action where one systematically takes the tension to zero, following the seminal analysis of \cite{Isberg:1993av}. When one identifies the metric by comparing with the Polyakov action \refb{weyl}, one finds that the determinant of $g^{\a\b}$ is zero. One can replace this degenerate metric density $T \sqrt{-g} g^{\alpha \beta}$ by a rank one matrix that can be written as $V^\alpha V^\beta$ where $V^\alpha$ is a vector density. The action in the $T \to 0$ limit, which we will refer to as the LST action after the authors \cite{Isberg:1993av}, becomes 
\be{lst}
S = \int d^2 \xi \,\ V^\alpha V^\beta \p_\alpha X^m \p_\beta X^n \eta_{mn}.
\ee
At this point, we can forget where the action \refb{lst} came from and treat it as the starting point of a theory of fundamentally tensionless strings. We will call this the intrinsic framework for this theory. The equations of motion as derived from \refb{lst} are: 
\be{eom} 
\p_\a(V^\a V^\b \p_\b X^\mu)=0, \quad V^\b\gamma_{\a \b}=0. 
\ee
where $\gamma_{\alpha \beta} = \p_\alpha X^m \p_\beta X^n \eta_{mn}$ is the induced metric on the worldsheet. The second equation in \refb{eom} indicates that $\gamma_{\a \b}$ is degenerate \cite{Isberg:1993av}. 

\bigskip

\noindent {\em{Symmetries of the worldsheet}} 

\medskip

\noindent Under a diffeomorphism $\xi^\a\rightarrow \xi^\a+\epsilon^\a$, the vector density $V^\a$ transforms as:
\be{} \delta V^\a=-V^\beta\p_\beta \epsilon^\a + \epsilon^\beta \p_\beta V^\a +\frac{1}{2}(\p_\beta \epsilon^\beta) V^\a. \ee
The action of the tensionless string is invariant under these worldsheet diffeomorphisms and hence we will fix a gauge: 
\be{v0}
V^\alpha = (1, 0).
\ee
There is a residual symmetry that is left over after this gauge fixing. The form of $\e^\a$ which leaves the gauge fixed action invariant is
\be{e-ur} 
\epsilon^\a=\{f'(\sigma)\tau+g(\sigma),f(\sigma)\}. 
\ee
For a function $F(\xi^a)$, the effect of such a transformation is given by:
\be{} \delta F=[f'(\sigma)\tau\p_\tau+f(\sigma)\p_\sigma+g(\sigma)\p_\tau]F =[L(f)+M(g)]F. \ee
Thus the generators of this residual gauge symmetry can be defined as: 
\begin{subequations}\label{LM}
\bea{} 
&& L(f)=f'(\sigma)\tau\p_\tau+f(\sigma)\p_\sigma=\sum_n a_n e^{in\sigma}(\p_\sigma+in\tau\p_\tau)=-i\sum_n a_n L_n, \\
&& M(g)=g(\sigma)\p_\tau=\sum_n b_n e^{in\sigma} \p_\tau=-i\sum_n b_n M_n 
\eea
\end{subequations}
where $f=\sum a_n e^{in\sigma},\ g=\sum b_n e^{in\sigma}$ have been expanded in Fourier modes. The algebra of the modes:
\be{bms} 
[L_m,L_n]=(m-n)L_{m+n}, \quad [L_m,M_n]=(m-n)M_{m+n}, \quad [M_m,M_n]=0. 
\ee
This is indeed the classical part of BMS$_3$ algebra \refb{BMS3}, which replaces the two copies of the Virasoro algebra on the tensile worldsheet as the residual gauge symmetry in the tensionless limit of bosonic string theory.   

\bigskip

\noindent {\em{Energy-momentum tensor}} 

\medskip

\noindent We consider an infinitesimal transformation of \refb{lst}: $\xi^\a\rightarrow\xi'^\a=\xi^\a+\delta\xi^\a$. The Noether current is given by $J^\a=T^\a_{\ \beta}\delta\xi^\beta$. The energy momentum (EM) tensor constructed from the above: 
\be{} 
T^\a_{\ \beta}=V^\a V^\rho \p_\rho X^\mu \p_\beta X_\mu-\frac{1}{2}V^\lambda V^\rho \p_\lambda X^\mu\p_\rho X_\mu \delta^\a_{\ \beta}. 
\ee
In the gauge $V^\a=(1,0)$, $\delta\xi^\a=(f'\tau+g,f)$ and the non-trivial components of $T^\a_{\ \beta}$ are
\be{} 
T^0_{\ 1} = \dot X\cdot X' \equiv T_1 (\s, \t),  \quad  T^0_{\ 0}=-T^1_{\ 1}=\frac{1}{2} \dot X^2 \equiv T_2 (\s, \t). 
\ee
The Noether current associated with the above transformation is: 
\be{}
 Q=\int d\sigma J^0 = \int d\sigma \left[T_1 f + T_2 (f'\tau+g) \right]. 
\ee
Expanding $f$ and $g$ in fourier modes as before we get:
\be{}  
Q=\sum_n a_n \int d\sigma \ (T_1+ in\tau T_2)e^{in\sigma}+\sum_n b_n \int d\sigma \ T_2 e^{in\sigma} = \sum_n a_n L_n +\sum_n b_n M_n. 
\ee
Thus we have:
\be{} 
L_n=\int d\sigma (T_1+in\tau T_2)e^{in\sigma}, \quad M_n=\int d\sigma\ T_2  \ e^{in\sigma}. 
\ee
We can invert this relation to find
\be{Tmode}
T_1 (\s, \t) = \frac{1}{2\pi}\sum_{n} (L_n - in\tau M_n) e^{-i n\s}, \quad T_2 (\s, \t)= \frac{1}{2\pi}\sum_{n} M_n e^{-in\s}.
\ee
These relations define the EM tensor for the underlying 2d field theory. 

\medskip

\noindent {\em{Mode Expansions}} 

\medskip

\noindent The equation of motion \refb{eom} in the \refb{v0} gauge assume a particularly simple form:
\be{xeom}
\ddot{X}^\mu=0.
\ee
The equations corresponding to $V^\a$, which are of course the two components of the energy momentum tensor that we have introduced above, become constraints:
\be{con}
\dot{X}\cdot X'=0=T_1, \quad \dot{X}^2=0=T_2. 
\ee
So the classical tensionless bosonic string is a system governed by the equation of motion \refb{xeom} subject to the constraints \refb{con}. 
Subject to closed string boundary conditions $X^\mu(\tau,\sigma)=X^\mu(\tau,\sigma+2\pi)$, the above EOM is solved by the following mode expansion:
\be{mode} 
X^{\mu}(\sigma,\tau)=x^{\mu}+\sqrt{\frac{c'}{2}}B^{\mu}_0\tau+\sqrt{\frac{c'}{2}}\sum_{n\neq0}\frac{i}{n} \left(A^{\mu}_n-in\tau B^{\mu}_n \right)e^{-in\sigma}. 
\ee
Using the mode expansion above on the equations of the constraints, and equating with the expansion of the EM tensor derived in \refb{Tmode}, we find:
\be{lmab} 
L_n= \frac{1}{2} \sum_{m} A_{- m}\cdot B_{m+n}, \quad M_n= \frac{1}{2} \sum_{m} B_{-m}\cdot B_{m+n}. 
\ee
Let's look at the algebra of these modes. The Possion brackets between $X$ and $P$ require
\be{AB} 
\{A^{\mu}_m,A^{\nu}_n\}_{P.B.}= \{B^{\mu}_m,B^{\nu}_n\}_{P.B.}=0, \quad \{A^{\mu}_m,B^{\nu}_n\}_{P.B.}= - 2 im\delta_{m+n}\eta^{\mu \nu}. 
\ee
It is important to stress that this is {\em{not}} the algebra of harmonic oscillator modes. The algebra of $\{A,B\}$ modes can be used to calculate those of $L_n, M_n$. This, as expected, leads to the BMS algebra \refb{bms} (upon quantisation). If we insist of a harmonic oscillator algebra, we need to define new modes $C, \C$: 
\be{CC}
C^{\mu}_n = \frac{1}{2}({A}^{\mu}_n+B^{\mu}_{n}), \quad \C^{\mu}_n =\frac{1}{2}(-{A}^{\mu}_{-n}+B^{\mu}_{-n}).
\ee 
Now, of course, the Poisson brackets take the canonical form: 
\be{4.22}
\{C^{\mu}_n, C^{\nu}_m \} = -i n \delta_{n+m, 0} \ \eta^{\mu \nu}, \quad  \{ \C^{\mu}_n,  \C^{\nu}_m \} = -i n \delta_{n+m, 0} \ \eta^{\mu \nu}, \quad \{C^{\mu}_n, \C^{\nu}_m \} = 0.
\ee
This algebra is clearly of the same form as in the usual tensile bosonic string oscillators $(\a, \tilde{\a})$. In this tensionless oscillator basis, we can alternatively write the mode expansion \refb{mode} which solves the equation of motion as
\be{}
X^\mu(\t,\s)=x^\mu+2\sqrt{\frac{c'}{2}}C^\mu_0\t+i\sqrt{\frac{c'}{2}}\sum_{n\neq 0} \frac{1}{n}\left[(C^\mu_n-\C^\mu_{-n})-in\t (C^\mu_n+\C^\mu_{-n})\right]e^{-in\s}.
\ee
It would be instructive to split this in the following way: 
\bes\bea{}
X^\mu_{``L"}&=&\frac{x^\mu}{2}+\sqrt{\frac{c'}{2}}C^\mu_0\t+i\sqrt{\frac{c'}{2}}\sum_{n\neq 0} \frac{1}{n}[C^\mu_n-in\t C^\mu_n]e^{-in\s} \\
X^\mu_{``R"}&=&\frac{x^\mu}{2}+\sqrt{\frac{c'}{2}}\C^\mu_0\t+i\sqrt{\frac{c'}{2}}\sum_{n\neq 0} \frac{1}{n}[\C^\mu_{n}-in\t \C^\mu_{n}]e^{in\s}.
\eea\ees
In the above, the zeroth mode of the oscillators are related to the string momentum: $C^\mu_0=\C^\mu_0=\sqrt{\frac{c'}{2}}k^\mu$. These structures imply that $C$ oscillators count the number of ``left'' modes and $\tilde C$ oscillators count the number of ``right'' modes, so that the original left-right splitting in the tensile string has a similar tensionless analogue.


\newpage

\noindent {\em{The Ultra-Relativistic limit}} 

\medskip

\noindent The tensionless limit is a limit where the length of the fundamental string goes to infinity. This can be viewed in terms of a limit on the co-ordinates of the worldsheet: $\s \to \infty$ keeping $\t$ fixed. Since we wish to impose closed string boundary conditions, it is easier to work with an equivalent limit
\be{URlim}
\s \to \s, \ \t \to \e \t, \ \e \to 0.
\ee
This is an {\em{ultra-relativistic}} (UR) limit on the worldsheet \footnote{The NR limit on the worldsheet can similarly be expressed as \be{NRlim}
\s \to\e \s, \ \t \to  \t, \ \e \to 0.
\ee See \cite{Bagchi:2015nca} for more details on this limit.}, as mentioned in the introduction, where the worldsheet speed of light goes to zero. The residual symmetry algebra on the worldsheet of the tensile bosonic closed string is given by two copies of the Virasoro algebra.
The symmetry in the tensionless limit manifests itself exactly as the UR contraction of the Virasoro generators (\ref{cont2}),
\be{vir2bms}
L_n= \L_n - \bL_{-n}, \ M_n = \e(\L_n + \bL_{-n}).
\ee
Starting with the Virasoro generators 
at the level of the algebra, the above UR limit on the two copies of the Virasoro results in the quantum version of the BMS$_3$ (with central terms turned on) as described in (\ref{BMS3}).

In terms of the mode expansions of the string, one can compare the tensile \refb{tensilemode} and the tensionless expansions \refb{mode} to easily come up with
\be{ABa}
A_n^{\mu} = \frac{1}{\sqrt{\e}} \left( \a_n^\mu - \tilde{\a}_{-n}^\mu \right), \quad B_n^{\mu} = {\sqrt{\e}} \left( \a_n^\mu + \tilde{\a}_{-n}^\mu \right). 
\ee
Here $\a_n$ and $\tilde{\a}_{n}$ are of course the modes of the usual tensile closed bosonic string. The relation \refb{ABa} also independently follows from the previous equations \refb{lmab} and \refb{vir2bms}. However, talking in terms of $C$ oscillators is much more instructive since we can define those in terms of the tensile ones by means of Bogoliubov transformations:
\be{c1}
C^\mu_n(\e) =\cosh \theta \ \a^\mu_n+\sinh \theta \ \ta^\mu_{-n}, \quad \C^\mu_n(\e) =\sinh \theta \ \a^\mu_{-n}+\cosh \theta \ \ta^\mu_{n}, 
\ee
where the relation between $\theta$ and $\e$ can be easily read off:
\be{}
\cosh\theta = \frac{1}{2} \left( \sqrt{\e} + \frac{1}{\sqrt{\e}} \right), \ \sinh\theta = \frac{1}{2} \left( \sqrt{\e} - \frac{1}{\sqrt{\e}} \right).
\ee
It is clear that at $\e=1$, the set of oscillators $\{C(\e), \C(\e)\}$ become the tensile $\a$ modes, and in the other extreme $\e\to 0$ takes them to the tensionless oscillators which we defined earlier in \refb{CC}. Hence the flow in $\e$ from 1 to 0 takes one systematically from tensile to tensionless strings.

\medskip

The UR limit also works on the EM tensors of the field theory in question. We can start with a relativistic CFT in 2 dimensions with EM tensors $T(z)$ and ${\bar{T}}(\z)$ and transform this from the plane to the cylinder:
\be{}
T_{cyl}(\w) = z^2 T(z) - \frac{c}{24} = \sum_n \L_n e^{-in\w} - \frac{c}{24}
\ee
where the central term comes from the piece in the transformation involving the Schwarzian derivative. There is a similar expression for $\bar{T}_{cyl}(\bw)$. We obtain the BMS versions of the EM tensors on taking the following linear combinations: 
\begin{subequations}
\bea{}
&& T_1(\s,\t)= T_{cyl}(\w) - \bar{T}_{cyl}(\bw) = \sum_n (L_n - in\t M_n) e^{-in\s} - \frac{c_L}{24} , \\ && T_2(\s,\t)= \e(T_{cyl}(\w) + \bar{T}_{cyl}(\bw))= \sum_n M_n e^{-in\s} - \frac{c_M}{24}. 
\eea
\end{subequations}
We can see that \refb{Tmode} is the version of the above equations without the central terms. 

\bigskip

\section{Imposing Quantum Constraints} 
\label{five}

Previously we have discussed the classical aspects of tensionless strings in some detail. In this section we are going to focus on the quantum aspects of it. We have explored the oscillator construction of the bosonic tensionless string and seen how the constraints can be expressed in terms of the oscillators. We have also looked at how the oscillators and constraints map to their tensile analogues. 
It is while quantising the string when we start to see interesting developments. Similar to the tensile case, we look at the method of covariant quantisation, where the oscillators are used to define a vacuum. We then build a Hilbert space from the vacuum. Applying the constraints on the Hilbert space gives us the physical string spectrum and its properties. Here, we shall see how we think out of the box to look at possible ways of imposing the constraints in the tensionless case. We narrow down the possibilities to three distinct cases and discuss their unique features in the succeeding sections. 

\subsection*{The principal philosophy of the paper} 
 
Let us consider that we have one generic constraint $F_n$ which is hermitian by definition ($F^\dagger_n=F_{-n}$). Classically we impose the constraint by simply saying $F_n=0$, $\forall\ n\in\mathbb{Z}$. When we quantise the system, we use the quantum version of such constraints to restrict the Hilbert space to filter out the physical spectrum. The most general way to impose this condition is demanding all the matrix elements of the constraint acting on physical states vanish, i.e. 
\be{Constraint}
\<phys'|F_n|phys\>=0\quad (n\in \mathbb{Z}).
\ee 
In the rest of the paper, we will refer to this condition as a {\textit{``sandwich''}} condition. For tensile string case, these constraints are left and right Virasoro generators: $F_n=(\mathcal{L}_{n},~ \mathcal{\bar L}_{n})$. Remember that in section \refb{F11} we have studied two consistent ways to impose the constraint condition consistently on a worldsheet, the first one being the conventional method of Virasoro highest weight representations:
\be{tensecond}
\mathcal{L}_{n} |phys\rangle= \mathcal{\bar L}_{n}|phys\rangle = 0\quad (n>0).
\ee
It should be noted that the sandwich conditions here work via the
\textit{right handed action} of the constraints. The other method is the case of the ``flipped'' vacuum where half of the conditions are that of the lowest weight, 
\be{ambicond}
\mathcal{L}_{n} |phys\rangle= \mathcal{\bar L}_{-n}|phys\rangle = 0\quad (n>0).
\ee
Notice in the case above, the anti-holomorphic constraints actually impose a \textit{left handed action} to satisfy the sandwich condition. 

For the tensionless case, the emergence of BMS$_3$ algebra makes the matters more conceptually difficult as there could be more possibilities to consistently define the string vacuum and physical states. We will see that this general sandwich condition, together with the property of hermiticity can be broken down into three distinct cases:
\begin{subequations}\label{Q1}
\bea{}
1.\ \ F_n|phys\>&=&0\quad (n>0), \\
2.\ \ F_n|phys\>&=&0\quad (n\neq 0), \\
3.\ \ F_n|phys\>&\neq& 0, \ \textit{but}\ \<phys'|F_n|phys\>=0.
\eea
\end{subequations}
Zero modes are not included here since one can always have an ordering ambiguity in those modes for which we need to consider $F_0=:F_0:-a_F$ in the above classification. The most comfortable and nice way is the first case because the physical states fall into the highest weight representation of the algebra (case 1), which is often the usual norm to study a quantum string theory with.
 
In the case of the BMS$_3$ algebra things are not that simple, and one needs to consider all the cases to understand the associated string spectrum. Here we have $F_n=(L_n, M_n)$ for which the above classification of conditions are possible. It seems that we could have nine possible combinations in total through which we can impose the constraint on the states. These are depicted below:
\bes \label{6.5} \bea{}
L_m|phys\>&=&0,\ (m>0),\ \left \{ \begin{matrix} M_n|phys\>=0,\ (n> 0) \\
M_n|phys\>=0,\ (n\neq 0) \\
M_n|phys\>\neq 0,\ (\forall\ n)\ \ \ \end{matrix} \right \}; \\
L_m|phys\>&=&0,\ (m\neq 0),\ \left \{ \begin{matrix} M_n|phys\>=0,\ (n> 0) \\
M_n|phys\>=0,\ (n\neq 0) \\
M_n|phys\>\neq 0,\ (\forall\ n)\ \ \ \end{matrix} \right \}; \\
L_m|phys\>&\neq&0,\ (\forall\ m),\ \ \ \left \{ \begin{matrix} M_n|phys\>=0,\ (n> 0) \\
M_n|phys\>=0,\ (n\neq 0) \\
M_n|phys\>\neq 0,\ (\forall\ n)\ \ \ \end{matrix} \right \}. 
\eea\ees
We must keep in mind that all of the above conditions give back the \textit{sandwich} condition \refb{Constraint}. We are going to discuss how choosing different vacuum configurations amounts to having the corresponding imposition of constraints. Let us remind the reader the relation of the constraints with the modes
\be{LM} L_n= \frac{1}{2} \sum_{m} A_{- m}\cdot B_{m+n}, \quad M_n= \frac{1}{2} \sum_{m} B_{-m}\cdot B_{m+n}. \ee
Reading off from \refb{AB} we write the following commutator brackets 
\be{Q4} 
[A^{\mu}_m,A^{\nu}_n]=[B^{\mu}_m,B^{\nu}_n]=0, \quad [A^{\mu}_m,B^{\nu}_n]=2m\delta_{m+n}\eta^{\mu \nu}. 
\ee
Evidently, using these brackets one also arrives at the same constraint algebra as in \refb{BMS3}. This particular construction is crucial for the constraint algebra to close. To impose the physical conditions on the Hilbert space we are going to make an assumption: The vacuum state \textit{must} be a physical state with respect to the constraints, at least for $n\neq 0$ i.e. $\<0|L_n|0\>=\<0|M_n|0\>=0,\ (n\neq 0)$. This is a strong condition since it accounts for any arbitrary $n$. For $n=0$, the ordering ambiguity plays a role in restricting the physical spectrum and we shall address this in more detail once we discuss each of the cases. 
 
Since both the constraints contain the oscillator $B_n$, we are going to impose the constraints starting with this oscillator and then derive the set of conditions on the vacuum for both $A_n$ and $B_n$ modes. 
First, let us see how we can define the vacuum state in the theory consequently from the set of conditions given in \refb{Q1}, if we are given one set of oscillators $B_n$, 
\begin{subequations}\label{Q2}
\bea{}
1.\ \ B_n|0\>&=&0\quad (n>0), \\
2.\ \ B_n|0\>&=&0\quad (n\neq0), \\
3.\ \ B_n|0\>&\neq& 0,\ \textit{but}\ \<0|B_n|0\>=0\quad (n\neq 0).
\eea
\end{subequations}
Out of the above three cases, imposing the third one is a bit tricky, but surprisingly it appears that we can interpret this last condition in \refb{Q2} by redefining the modes in terms of linear combinations $B_n=C_n+\C_{-n}$ of two oscillators \refb{CC}, where the positive modes of $C_n$ and $\C_n$ annihilate the associated vacuum. We can immediately see that the harmonic oscillator basis of the tensionless string modes \refb{CC} is naturally useful in this sense. Consequently, from \refb{Q2} we can derive nine sub-cases, if we apply the same scenarios for $A_n$ oscillators. These nine cases are connected to the nine possible combinations of $L_n$, $M_n$ given in \refb{6.5}. This connection is illustrated in table \refb{t1} below.
\begin{center}
\begin{table}[h!]
\begin{tabular} {|c|c|c|c|c|c|}
\hline
Label & Constraint & Oscillator & Condition & Vacuum \\
\hline \hline
(1,1) & $L_m |\varphi\> = 0, M_n |\varphi\> = 0$ & $A_m|0\>=0, B_n|0\>=0$ & $m>0, n>0$ &Flipped: $|0\>_A$ \\
\hline
(1,2) & $L_m |\varphi\> = 0, M_n |\varphi\> = 0$ & $A_m|0\>=0, B_n|0\>=0$ & $m>0, n\neq 0$ & Inconsistent \\
\hline
(1,3) & $L_m |\varphi\> = 0, M_n |\varphi\> \neq 0$ & $A_m|0\>=0, B_n|0\>\neq0$ & $m>0, n\in \mathbb{Z}$ & Inconsistent \\
\hline 
(2,1) & $L_m |\varphi\> = 0, M_n |\varphi\> = 0$ & $A_m|0\>=0, B_n|0\>\neq0$ & $m\neq 0, n>0$& Inconsistent \\
\hline
(2,2) & $L_m |\varphi\> = 0, M_n |\varphi\> = 0$ & $A_m|0\>=0, B_n|0\>=0$ & $m\neq 0, n\neq 0$ & Induced: $|0,0^\mu\>_I$ \\
\hline
(2,3) & $L_m |\varphi\> = 0, M_n |\varphi\> \neq 0$ & $A_m|0\>=0, B_n|0\>\neq0$ & $m\neq 0, n\in \mathbb{Z}$ & Inconsistent \\
\hline 
(3,1) & $L_m |\varphi\> \neq 0, M_n |\varphi\> = 0$ & $A_m|0\>\neq 0, B_n|0\>=0$ & $m\in \mathbb{Z}, n>0$& Inconsistent \\
\hline
(3,2) & $L_m |\varphi\> \neq 0, M_n |\varphi\> = 0$ & $A_m|0\>\neq 0, B_n|0\>=0$ & $m\in \mathbb{Z}, n\neq 0$& Induced: $|0,k^\mu\>_I$ \\
\hline
(3,3) & $L_m |\varphi\> \neq 0, M_n |\varphi\> \neq 0$ & $A_m|0\>\neq 0, B_n|0\>\neq0$ & $m\in \mathbb{Z}, n\in \mathbb{Z}$& Oscillator: $|0\>_C $ \\
\hline
\end{tabular}
\caption{Quantum constraints on a physical state $|\varphi\>$ leading to different vacua.}
\label{t1}
\end{table}
\end{center}

In the table mentioned, the different implementations are labeled by $(l,m)$ where denoting the type of condition listed in \refb{Q1} or \refb{Q2}. The highest weight representations are be labeled by $(1,1)$. However as mentioned in the table, most of the nine combinations gives us inconsistencies or redundancies, since the two constraints are not completely arbitrary but are related through the structure of BMS algebra. A detailed calculation of why the other constraints do not hold is carried out in Appendix \ref{ApB}. We list the three main interesting cases for imposing the physical conditions:
\begin{subequations}\label{Q5}
\bea{}
i.\ \ L_n|phys\>&=&M_n|phys\>=0\quad (n>0), \\
\label{5.89b} ii.\ \ L_n|phys\>&\neq&0,\quad M_n|phys\>=0\quad (n\neq0), \\
iii.\ \ L_n|phys\>&\neq& 0,\quad M_n|phys\>\neq0.
\eea\ees
Reading off from the table, these gives us the three distinct vacua (cases (3,3),(3,2) and (1,1)) which we name as the oscillator, induced and flipped vacua. Case (2,2) is also a valid choice and turns out to be a special case of the induced vacuum where the momentum is zero. Now, in what follows, we delve into a detailed discussion for each of these vacua.

\bigskip

\section{The Oscillator Vacuum}\label{osc}
We begin by considering the tensionless string action \refb{lst} as fundamental. This means that we will attempt to quantize the theory defined by this action without recourse to any parent theory. The equation of motion \refb{eom} in the $V^\a = (1,0)$ gauge, as seen before, is solved by the mode expansion \refb{mode}. There the oscillator modes $(A, B)$, as stressed before, don't follow a harmonic oscillator algebra, but instead the commutation relations \refb{AB}. To work in a familar oscillator basis we shall work with the $C$ oscillators. Upon canonical quantization, the Poisson brackets \refb{4.22} transform to 
\be{c2}
[C^\mu_m,C^\nu_n]=[\C^\mu_m,\C^\nu_n]=m\eta^{\mu\nu}\delta_{m+n}.
\ee
The {\em{oscillator vacuum}} $\zc$ is defined by the following
\be{c3}
C^\mu_n\zc=\C^\mu_n\zc=0\quad\forall\ n>0.
\ee
We call this vacuum the oscillator vacuum since the definition exactly mirrors that of a tensile string vacuum. 

\smallskip

\noindent {\em{The Hilbert space: A first look.}}
\smallskip
\smallskip

In the following discussion, we would however be more interested to account for the string states in a more intrinsic way.
The relations \refb{c2} allows us to create states by acting negative modes of $C$ and $\C$ on the vacuum. Thus a Hilbert space can be constructed in the following way:
\bea{c5}
\mbox{Level}\ 0:&&\ |0,k^\mu\>_c \nonumber \\
\mbox{Level}\ 1:&& \ C^\mu_{-1}|0,k^\mu\>_c; \quad \C^\mu_{-1}|0,k^\mu\>_c \nonumber \\
\mbox{Level}\ 2:&& \ C^\mu_{-2}|0,k^\mu\>_c; \quad C^\mu_{-1}C^\nu_{-1}|0,k^\mu\>_c; \nonumber \\
&&\ C^\mu_{-1}\C^\nu_{-1}|0,k^\mu\>_c;  \quad \C^\mu_{-1}\C^\nu_{-1}|0,k^\mu\>_c; \quad \C^\mu_{-2}|0,k^\mu\>_c  \nonumber \\
\hdots&& \ \hdots\quad\hdots\quad\hdots\quad\hdots \quad\hdots \quad\hdots \quad\hdots \quad\hdots \nonumber \\
\mbox{Level}\ n:&& \ C^\mu_{-n}|0,k^\mu\>_c; \quad C^\mu_{-1}C^\nu_{-n+1}|0,k^\mu\>_c;\quad\hdots\quad\C^\mu_{-n}|0,k^\mu\>_c .
\eea
A generic state $|\psi\rangle$ with level \textbf{(r+s)} is constructed as
\be{c6}
|\psi\rangle=|\textbf{r,s}\rangle=\sum_{j} \rho_j\odot\Big[\Big(C^{(a_1)}_{-{m_1}}...C^{(a_p)}_{-{m_p}}\Big)\Big(\C^{(b_1)}_{-{n_1}}...\C^{(b_q)}_{-{n_q}}\Big)\Big]_j|0,k^\mu\>_c, 
\ee
where $(a_i)$ and $(b_i)$ denote the power of the $C_{-m_i}$ and $C_{-n_j}$ oscillators. Spacetime indices have been suppressed in the above for simplicity. 
\be{} 
r=\sum_{i=1}^pa_im_i, \quad s=\sum_{i=1}^qb_in_i, \quad (r, s \geq 0)
\ee 
count the total number of $C$ and $\C$ modes respectively. In \refb{c6}, $\rho_j$ are polarisation tensors with the appropriate index structure. It is easy to see that this state has mass squared equal to $-k^2$. 

\smallskip

Let us now consider the inner products of the states in the Hilbert space. Firstly, we consider the vacuum to be normalised i.e. 
\be{}
\<0,k^\mu|0,k'^\mu\>=\delta_{kk'}.
\ee 
Also, the excited states at different levels are orthogonal to each other. 
An obvious fact arising from the orthogonality condition is that the inner product between a level matched state and a non-level matched state is zero. This will be important later in our analysis. We will now focus on the implementation of worldsheet constraints to understand the physical states in the Hilbert space. 

\subsection{Action of constraints}
Now we analyse in detail the action of constraints. To write the constraints, we use the definition in (\ref{CC}) and put in them back into (\ref{lmab}). The constraints then can be expressed in terms of the $C$ oscillators
\begin{subequations}\label{c7}
\bea{}
L_n&=&\frac{1}{2} \sum_m \big[ C_{-m}\cdot C_{m+n} - \C_{-m}\cdot\C_{m-n} \big],\\
M_n&=&\frac{1}{2} \sum_m \big[ C_{-m}\cdot C_{m+n} + \C_{-m}\cdot\C_{m-n} + 2C_{-m}\cdot\C_{-m-n} \big]. 
\eea 
\end{subequations}
For $n\neq0$ this can be expressed in terms of new operators $\J_n$, $\bJ_n$ and $\K_n$ such that :
\be{c8}
L_n=\J_n-\bJ_{-n}, \quad M_n=\J_n+\bJ_{-n}+2\K_{n}.
\ee
where we define the new operators as,
\bea{c9}
\J_n=\frac{1}{2} \sum_m C_{-m}\cdot C_{m+n}, \ 
\bJ_n=\frac{1}{2} \sum_m \C_{-m}\cdot\C_{m+n}, \
\K_n=\frac{1}{2} \sum_m C_{-m}\cdot\C_{-m-n}. 
\eea
For the normal ordered zero modes, it is more convenient to write them as  
\be{c10}
L_0=\N-\bar{\N}, \quad
M_0=c'k^2+\N+\bar{\N}+X+X^\dagger,
\ee
where, 
\be{c11}
\N=\sum_{m>0} C_{-m}\cdot C_{m}; \quad \bar{\N}=\sum_{m>0} \C_{-m}\cdot \C_{m}; \quad X=\sum_{m>0} C_m\cdot\C_m.
\ee 
Here $\N$ and $\tilde{\N}$ are the number operators for either set of oscillators, while $X$ is a sum of annihilation operators that couples the two sets. Action-wise, we see that the number operators in $M_0$ are pieces that give an eigenvalue when it acts on the states, i.e. a diagonal part, which we denote as $H_0$. There is the other coupled piece that instead gives rise to  a bunch of other states on action, we call that as $Y_0$. 
\be{c12}
M_0=H_0+Y_0, \quad
\text{where }H_0=c'k^2+\N+\tilde{\N},\ Y_0=X+X^\dagger.
\ee
The constraint algebra can be checked to be the BMS algebra \refb{BMS3}. 

\medskip

We now wish to find the central extensions of the BMS$_3$ in this vacuum. By considering Jacobi identities and the inner product of commutators $[L_2, L_{-2}]$ and $[L_2, M_{-2}]$ between the vacuum with zero momentum $|0,0^\mu\>_c$, the values of the central charges can be calculated explicitly in terms of the spacetime dimensions $D$ (See \ref{ApA1}). We find these charges to be
 \be{c14} c_L=0\quad \mbox{and} \quad c_M=2D. \ee
Let us now consider the action of the constraint operators on a general state $|\textbf{r,s}\>$, where in general both levels do not have to be same. $\J_n$ lowers the level of $C$ oscillators by $n$ unless it becomes zero. $\bJ_n$ does the same thing for $\C$ oscillators, 
\be{c15}
\J_n|\textbf{r,s}\>=|\textbf{r-n,s}\>, \quad \bJ_n|\textbf{r,s}\>=|\textbf{r,s-n}\>\quad\forall\ n\neq0. 
\ee
The action of $\K_n$ on $|\textbf{r,s}\>$ affects both levels of the state and is given by the following:
\begin{subequations}\label{c16}
\bea{}
\K_n|\textbf{r,s}\>&=&\sum_m |\textbf{r-m+n,s-m}\>+\sum_m|\textbf{r+m,s+m+n}\>\quad\forall\ n>0, \\
&=&\sum_m|\textbf{r-m,s-m+n}\>+\sum_m|\textbf{r+m+n,s+m}\>\quad\forall\ n<0. 
\eea
\end{subequations}
We can collect these expressions to obtain the combined operation of $L_n$ and $M_n$'s on the state $|\textbf{r,s}\>$:
\begin{subequations}\label{c17}
\bea{}
L_n|\textbf{r,s}\>&=&|\textbf{r-n,s}\>-|\textbf{r,s+n}\> \quad (n \neq 0), \\
M_n|\textbf{r,s}\>&=&|\textbf{r-n,s}\>+|\textbf{r,s+n}\>+\sum_{m>0}\Big[|\textbf{r-m+n,s-m}\>+|\textbf{r+m,s+m+n}\>\Big](n>0)\nonumber \\
&=&|\textbf{r-n,s}\>+|\textbf{r,s+n}\>+\sum_{m>0}\Big[|\textbf{r-m,s-m-n}\>+|\textbf{r+m-n,s+m}\>\Big](n<0).
\eea
\end{subequations}
The action of zero modes can also be calculated quite easily:
\begin{subequations}\label{c18}
\bea{}
L_0|\textbf{r,s}\>&=&(r-s)|\textbf{r,s}\>, \\
M_0|\textbf{r,s}\>&=&(c'k^2+r+s)|\textbf{r,s}\>+\sum_{m>0}\Big[|\textbf{r-m,s-m}\>+|\textbf{r+m,s+m}\>\Big].
\eea
\end{subequations}
The most important point to note is that if we start with a level matched state $|\phi_r\>=|\textbf{r,r}\>$, then the action of $L_n$ or $M_n$ on $|\phi_r\>$ will give us combinations of a number of states $\sim\sum_m|\textbf{a}_m,\textbf{b}_m\>$ that are not level matched at all $(a_m\neq b_m).$ 

\medskip

Let us pause to emphasise the importance of the above calculation. It clearly shows that the right hand action of constraints on a state isn't enough to probe this theory, even if the oscillator structure is preserved. Secondly, from an operator formalism point of view that when the constraint operators act on a physical state, they spew out an accumulation of infinite number of unphysical excitations. Interestingly, these unphysical excitations stem out from the $C,\C$ coupled term in the Hamiltonian of the theory. One may speculate that at the worldsheet level, the coupling of these oscillators themselves are unphysical as in some sense they are causally disconnected, leading to the non-trivial structure of physical constraints on the string. However, we will leave such ideas for discussion elsewhere, and impose the generalized physical conditions to find more about the spectrum.

\subsection{Imposing physical conditions}
As we have stressed in Sec 3, the quantum imposition of constraints lead us to the sandwich conditions \refb{Constraint}. However $L_0$ and $M_0$ have normal ordering ambiguities, so the general physical state conditions are generalised to:
\be{c19}
\<phys'|(L_n-a_L\delta_{n,0})|phys\>=0, \quad \<phys'|(M_n-a_M\delta_{n,0})|phys\>=0.
\ee
The standard way of imposing the constraints by imposing the highest weight conditions does not work in this case. This can be seen by looking at the eqns 
\refb{c7}. In $M_n$ there is the presence of the operator $\K_n$, which acting on physical states creates an infinite tower of excited states. We also see that $L_n$ or $M_n$ with $n>0$ \textit{does not} annihilate the vacuum \refb{c17}. It is easy to see that if we demand the right hand action of $L_n$ or $M_n$ on the vacuum $|0\>_C$ gives zero, it would give us a trivial result that all higher excitations on the vacuum will be zero. In order to obtain non-trivial states on this vacuum, one has to take the sandwich conditions \refb{c19} as guiding principle to find physical states. Let us see how this is applicable for the following situations: 

\paragraph{Case I : The Vacuum:}

Let us first consider $|phys\>=|phys'\>=|0,k_0^\mu\>_c$.  
It is easy to check that the physical conditions are trivially satisfied for $n\neq0$. For the zero modes we get
\be{c20}
{}_c\<0,k_0^\mu|L_0|0,k_0^\mu\>_c=\N-\bar{\N}=a_L=0, \quad {}_c\<0,k_0^\mu|M_0|0,k_0^\mu\>_c=c'k_0^2=a_M.
\ee
We see that the vacuum $|0,k_0^\mu\>_c$ is a physical state  with the mass shell condition being $m^2=-\frac{a_M}{c'}$, provided we have $a_L=0$ since $\N=\bar{\N}=0$.

\paragraph{Case II : Level matched states:}
Now let us take another state $|\Phi\>=|\textbf{r,r}\>$. Then the action of $L_n$ is given by
\bea{c21} 
L_n|\textbf{r,r}\>&=&|\textbf{s}_{-},\textbf{r}\>-|\textbf{r},\textbf{s}_{+}\>. 
\eea
where $s_{\pm}=r\pm n$. Curiously, the resultant state is clearly a non-level matched state for $n\neq0$. The same can be checked for the action of $M_n$. The inner product of this state with any other level matched state will be zero because of orthogonality. This gives us the following relations 
\be{c22}
\<\textbf{r,r}|L_n|\textbf{s,s}\>=0, \quad \<\textbf{r,r}|M_n|\textbf{s,s}\>=0. 
\ee
Recalling \refb{c18}, the $L_0$ condition simply gives us $L_0|\textbf{r,r}\>=0$, while for action of $M_0$ we can write  
\bea{c23}
M_0|\textbf{r,r}\>&=&(c'k^2+2r)|\textbf{r,r}\>+\sum_{m\neq0}|\textbf{r+m,r+m}\>.
\eea
As we explained before, the resultant sum of states are still individually level matched. Now let us consider the inner product with another level matched state $|\textbf{s,s}\>$, 
\bea{c24}
\<\textbf{s,s}|M_0|\textbf{r,r}\>&=&(c'k^2+2r)\<\textbf{s,s}|\textbf{r,r}\>+\sum_{m\neq0}\<\textbf{s,s}|\textbf{r+m,r+m}\>=a_M\<\textbf{s,s}|\textbf{r,r}\>
\eea
If $r=s$ then only the first term survives, else for $s=r\pm m$, the equivalent term inside the summation will survive. We will proceed to find out the mass shell condition for these states soon. We see from above that all level matched states $|\Phi\>$ obey the conditions: 
\be{c25}
\<\Phi'|L_n|\Phi\>=0,\quad \<\Phi'|M_n|\Phi\>=a_M\delta_{n,0}\<\Phi'|\Phi\>. 
\ee
The above relations are exactly the physical state conditions \refb{c19} with $a_L=0$. This nicely realises our framework for imposing the constraint conditions. 

\paragraph{Case III : Non level matched states:}
Let $|\Phi\>=|\textbf{r,s}\>$ such that $r-n=r'$ and $s+n=s'$  ($r,s>0$ and $n\neq0$). Then $L_n$ acting on $|\textbf{r,s}\>$ gives:
\bea{c26} L_n|\textbf{r,s}\>&=&|\textbf{r-n,s}\>-|\textbf{r,s+n}\>\nonumber \\
&=&|\textbf{r}',\textbf{s}\>-|\textbf{r},\textbf{s}'\>
\eea
 Then clearly we can find the following inner products:
\begin{subequations}\label{c27}
\bea{} \<\textbf{r}',\textbf{s}|L_n|\textbf{r,s}\>&=&\<\textbf{r}',\textbf{s}|\textbf{r}',\textbf{s}\>\neq0, \\
\<\textbf{r},\textbf{s}'|L_n|\textbf{r,s}\>&=&-\<\textbf{r},\textbf{s}'|\textbf{r},\textbf{s}'\>\neq0,
\eea
\end{subequations}
which violate physical state conditions. Using \refb{c17} similar relations can be derived for $M_n$ s with $n\neq0$.
 We can then say, there will be at least one value of $n$ where the physical conditions \refb{c19} are violated for a non level matched state. Furthermore, the action of $L_0$ will give us 
\bea{c28} 
L_0|\textbf{r,s}\>=(r-s)|\textbf{r,s}\>&\neq&0.
\eea
For non level matched states to be physical, it would mean that the ambiguity $a_L=r-s$ is not a fixed number, and since $a_L\neq0$, the vacuum is not a physical state.  We will demand for the theories which we will analyse, the respective vacua would be a physical state. This is a demand that is justified by looking at usual bosonic string theory{\footnote{One could obviously argue that for superstring theories, the GSO projection projects out the negative mass vacuum. But this is a condition that is imposed on top of the usual Super-Virasoro constraints.}}. Therefore, we conclude from the above discussion that {\textit{only level matched states can be physical states}} of the tensionless string defined on the oscillator vacuum. 

\subsection{Analysis of the physical spectrum}

We have seen that the level matched states trivially satisfy the $L_n$, $M_n$ ($n\neq0$) and $L_0$ conditions. The $M_0$ condition, on the other hand, help us to understand the mass spectrum of the physical states. In this section we are going to see how we get a massive spectrum and another massless spectrum of states by carefully analysing the constraints. Our focus is, of course, on level matched states from now on, as these are the only physical ones.

\subsubsection*{I. Massive states:}

Let us take a closer look at the $M_0$ condition for physical states. From \refb{c12}, we can split $M_0$ into ``good'' (number operators) and ``bad'' (coupled mode) parts: 
\begin{subequations}\label{c29}\bea{} 
\<phys'|M_0|phys\>&=&a_M\<phys'|phys\>,\\
\implies
\<phys'|H_0|phys\>&+&\<phys'|Y_0|phys\>=a_M\<phys'|phys\>.
 \eea\end{subequations}
Now, the action of $Y_0$ acting on the level matched physical state is given by the last term of \refb{c18}. Using the result, it can be checked that 
\be{}\<\textbf{n,n}|Y_0|\textbf{n,n}\>=0
\ee 
due to a mismatch of levels on both sides. On the other case, where the states are of different levels $\textbf{n}'\neq\textbf{n}$, 
\be{} \<\textbf{n}',\textbf{n}'|Y_0|\textbf{n,n}\>\sim\delta_{kk'}.
\ee 
This is equal to zero only if the momentum of different levels are not equal. We will now carefully analyse this condition. At first we address the case where $k\neq k'$. In the next subsection, we will look at $k=k'$, which would lead to massless states. For these states where $k\neq k'$, we can reduce the $M_0$ condition to just the sandwich action of $H_0$,
\be{c30} 
\<\textbf{n}',\textbf{n}'|H_0|\textbf{n,n}\>=a_M\<\textbf{n}',\textbf{n}'|\textbf{n,n}\>.
 \ee
Since $H_0$ is a ``good'' operator with well defined eigenvalues and eigenvectors, we can safely impose the right-action on the states:
\bea{c31}
H_0|\textbf{n,n}\>&=&a_M|\textbf{n,n}\>. 
 \eea
The case where $k=k'$, is discussed in the next section. Using the form of $H_0$ from (\ref{c12}), this condition allows us to define the mass of these states in terms of the ambiguity $a_M$,
 \bea{c32}
m^2|\textbf{n,n}\>&=&\frac{1}{c'}(2n-a_M)|\textbf{n,n}\>.
 \eea
As discussed above, other than the $M_0$ constraint, the other conditions doesn't really give us any new information regarding the states, just that these states have to be level matched. We have also seen that the simple right hand side action of the constraints on the states doesn't work. We shall see how we can impose the constraint in its sandwich form but with a minor additional assumption, which will help us to get meaningful physical states in the Hilbert space.
Let us consider the $L_n$ condition first. If we recall \refb{c8}, we can write 
\bea{c33} 
\<phys'|L_n|phys\>&=&\<phys'|\J_n|phys\>-\<phys'|\bJ_{-n}|phys\>=0\quad \forall\ n\neq 0. 
\eea
If we choose $|phys'\>$ as the physical vacuum, then from the form of $\mathcal{J}_n$'s make sure that this condition reduces to
\bea{c34} 
\<0,k^\mu_0|\J_n|phys\>&=& \<0,k^\mu_0|\bJ_n|phys\>=0\quad \forall\ n>0. 
\eea
The $M_n$ condition also gives the same result. Also, if we take a level matched state $|\Phi\>=|\textbf{m,m}\>$, then $\J_n|\textbf{m,m}\>$ is trivially zero for $n>m$. For $n\leq m$, we cannot deduce any meaningful condition, as whatever state forms on the right hand side, is orthogonal to the vacuum. Let us see what happens if we put a restriction on the states in terms of right hand actions:
\bea{c35} \J_n|phys\>&=& \bJ_n|phys\>=0\quad \forall\ n>0. \eea
The sandwich condition is then satisfied by the following way:
\bea{c36} \<phys'|L_n|phys\>&=&\<phys'|\overrightarrow{\J_n|phys}\>-\<\overleftarrow{phys'|\bJ_{-n}}|phys\>=0\quad \forall\ n>0, \nonumber \\
&=&\<\overleftarrow{phys'|\J_n}|phys\>-\<phys'|\overrightarrow{\bJ_{-n}|phys}\>=0\quad \forall\ n<0, \eea
while $L_n|\Phi\>$ is not necessarily zero. It is interesting to notice the similarity of $\J_n$ and $\bJ_n$ to the Virasoro constraints $\ML_n$ and $\bL_n$ in the tensile string theory and it apparently feels like we are going back to two copies of Virasoro. However that is not quite the case here, since the action of zero modes are defined in a different way \refb{c10}. 

\medskip

Let us illustrate an example by considering the state $|\textbf{1,1}\>=\rho_\mn C^\mu_{-1}\C^\nu_{-1}|0,k^\mu\>_c$. By demanding $\J_1|\textbf{1,1}\>=\bJ_1|\textbf{1,1}\>=0$, we would obtain $k^\mu\rho_\mn=k^\nu\rho_\mn=0$. 
In order to project out the negative norm states we must have 
\be{c37} \rho^2 \geq 0 \rightarrow k^2\leq 0. \ee
One can extend this analysis to higher levels, and obtain conditions relating the momentum and the polarisation tensor for which we have no negative norm states. 

\subsubsection*{II. Massless states:}

The reduction of the constraint conditions discussed in the previous section are valid only for physical states that have different momentum at different levels. In general there could be two states at different level built on the same vacuum $|0,k^\mu\>_c$. Then we do not have the freedom of setting the sandwich of the ``bad'' operator $Y_0$ to zero and the total physical condition can't be reduced to an analogue of (\ref{c31}). 

\medskip

For example, let us have a physical state as a linear combination of level matched states, $|\Phi\>=\lambda_1|\textbf{r,r}\>+\lambda_2|\textbf{s,s}\>$ and $|\Phi'\>=|\textbf{r,r}\>$ such that $r\neq s$. Now if we apply the $M_0$ condition, we get
\begin{subequations}\label{c38}
\bea{} 
\lambda_1\<\textbf{r,r}|M_0|\textbf{r,r}\>+\lambda_2\<\textbf{r,r}|M_0|\textbf{s,s}\>=\lambda_1a_M\<\textbf{r,r}|\textbf{r,r}\>, \\
\implies\lambda_1(c'k^2+2r)\delta_{kk'}+\lambda_2\sum_m\<\textbf{r,r}|\textbf{s+m,s+m}\>\delta_{r,s+m}=\lambda_1a_M\delta_{kk'}.
 \eea
 \end{subequations}
Therefore we see that there is a non-trivial contribution from both $H_0$ and $Y_0$ parts of $M_0$ for $k=k'$. An important question is how we can build physical level matched states using the generators $L$ and $M$'s in this case. In case of closed tensile strings we needed states $\sim\ML_{-n}\bar{\ML}_{-n}\za$ to construct level matched states. For our case, we can construct level matched states using some specific combinations of $L$ and $M$s such that it is annihilated by $L_0$. A problem with the states built using $M$'s is that these states run up to infinite levels due to the non-trivial action, therefore the simplest level matched state with a definite momentum $k$ that we can build on the vacuum is $|P_1\>=L_{1}L_{-1}\zc$. In terms of oscillators this has the form,
\bea{c39}
|P_1\>=L_1L_{-1}|0,k^\mu\>_c&=&c'k^2|0,k^\mu\>_c-c'k_\mu k_\nu C^\mu_{-1}\C^\nu_{-1}|0,k^\mu\>_c.
\eea
An interesting thing to note here is that this state is exactly equal to $L_{-1}L_{1}\zc$. This is true for any state of the form $L_nL_{-n}\zc$, due to the fact that $L_0$ acts on any level matched state to give zero and the central charge $c_L=0$. This state $|P_1\>$ is clearly a linear combination of $|\textbf{0,0}\>$ and $|\textbf{1,1}\>$ states as we explained above. For this state to be physical we now need to reaffirm the viability of  sandwich conditions,
\begin{subequations}\label{c40}
\bea{}
\<0,k_0^\mu|(M_0-a_M)|P_1\>&=&0, \\
\<P_1|(M_0-a_M)|P_1\>&=&0. 
\eea
\end{subequations}
Expanding these gives us the following results:
\begin{subequations}\label{c41}
\bea{}
c'k^2(c'k^2-1-a_M)\delta_{kk_0}&=&0, \\
2(c'k^2)^2(c'k^2-a_M)&=&0. 
\eea
\end{subequations}
Looking closely at the conditions, we find for $k\neq k_0$, we get either $k^2=0$ or $c'k^2=a_M$ from the second equation, but the latter can't be possible since action of $M_0$ dictates $c'k_0^2=a_M$ from (\ref{c20}). On the other hand, if $k=k_0$, then we get that $c'k_0^2=0=a_M$. This tells us that we must have $k=k_0$ and the ordering ambiguity in $M_0$ is fixed to be 
\be{c42} a_M=0. \ee 
This in turn means that {\textit{a physical ground state must have zero mass}}. 

It can also be checked that the norm of $|P_1\>$ is zero. In general if we apply the the $M_0$ condition to level matched states that are built on $|0,k^\mu_0\>_c$, i.e. they are level matched massless states, we can derive the relations
\begin{subequations}\label{c43} \bea{}
\<\textbf{n}',\textbf{n}'|M_0|\textbf{n,n}\>&\sim&f[\rho',\rho]=0, \\
\<\textbf{n,n}|M_0|\textbf{n,n}\>&\Rightarrow&(c'k^2_0+2n-a_M)||\textbf{n,n}||^2=0.
\eea \end{subequations}
Here $f[\rho',\rho]$ is a function relating the porlarisation tensors. Since we know that $c'k^2_0=a_M=0$ for these states, we must have the norm of these states $||\textbf{n,n}||^2=0$, i.e. all of them are null states. These level matched states are built on $|0,k^\mu_0\>$, therefore their masses are given by 
\be{c43a} m^2|\textbf{n,n}\>=m^2\Big(C^{\mu_1}_{-{l_1}}...C^{\mu_p}_{-{l_p}}\Big)\Big(\C^{\nu_1}_{-{m_1}}...\C^{\nu_q}_{-{m_q}}\Big)|0,k_0^\mu\>_c=0, \ee
where $\sum_i l_i=\sum_i m_i=n$. In summary,  we have discussed a clear classification of the tensionless spectrum built on the vacuum $\zc$. The vacuum state itself is physical if it is massless. There exist excited states that can be either massless null states, or massive states. For the latter ones, if $a_M=0$ the masses of excited states in \refb{c32} are always positive and the spectrum is classified by
\be{c44} m^2|\textbf{n,n}\>=\frac{2n}{c'}|\textbf{n,n}\>. \ee
This means $k^2<0$ and from \refb{c37} we can conclude that the states have positive norm. We can now see that by deviating from the conventional method of imposing constraints only the right side for positive modes, we get novel results simply by analyisng carefully the sandwich conditions of the constraints, which are more fundamental by nature. One of the most important result is the physical vacuum is massless: \textit{there is no tachyon in the bosonic tensionless theory over the oscillator vacuum.}

\medskip

The tensionless string on this vacuum clearly seems to have a memory of its parent oscillator structure. The oscillators $C$ and $\C$ arise from the evolution of the tensile oscillators $\a$ and $\ta$ in the limit \refb{c1}. The canonical relations, the level matching conditions for the string are preserved. Moreover, the mass formula in equation \refb{c44} is almost analogous to the formula for the tensile string except for the value of the ordering ambiguity and the string length $\a'$ replaced by the parameter $c'$. On the other hand we also have the tower of massless higher spin states (\ref{c43a}) that obey the physical conditions. Our discussion shows these are also the null states of the theory. In \cite{Bagchi:2015nca} such a hint of massless tower of states was uncovered for these tensionless strings in the sense that all masses vanish when one takes the $\a'\to \infty$ limit. In the present work, our intrinsic analysis solidifies this fact and we can further probe into a hitherto unknown massive sector of the theory.
\subsection{Critical dimensions?}

We have seen that it is difficult to construct physical states at a fixed level using the generators. These states certainly do not form a highest weight representation of the BMS Algebra. If we simply  had highest weight at our disposal, then we could have considered $|\phi\>=L_{-n}\zc$ or $|\phi'\>=M_{-n}\zc$ to be a physical state for $n>0$, and impose $L_n |\phi\>=L_n |\phi'\>=0$. This would mean that $|\phi\>$,$|\phi'\>$  are null physical states or spurious states, that exist for a particular value of the ambiguities and the central charges. These values can be calculated if we could set the right hand side of the following equations to zero. 
\begin{subequations}\label{c45} \bea{} \<\phi|\phi\>&=&{}_c\<0|L_n L_{-n}\zc \cr 
&=&{}_c\<0|\Big[2nL_0+\frac{c_L}{12}n(n^2-1)\Big]\zc-{}_c\<0|L_{-n} L_n\zc\quad\forall\ n>0, \\
\<\phi|\phi'\>&=& {}_c\<0|L_n M_{-n}\zc \cr
&=&{}_c\<0|\Big[2nM_0+\frac{c_M}{12}n(n^2-1)\Big]\zc-{}_c\<0|M_{-n} L_n\zc\quad\forall\ n>0. \eea \end{subequations} 
However, we cannot do so since the constraints does not act on the right hand side. The null states also can not be defined in this manner. In fact we have already seen that $L_nL_{-n}\zc\neq 0$, but instead generates combinations of level matched states. The central charge has also been calculated as $c_L=0$ for level matched states to exist in the physical state space. For our case, $|\phi\>$ or $|\phi'\>$ are not physical since they are not level matched, and the sandwich conditions \refb{c19} are not applicable. While we can show that the other central charge is related to the spacetime dimension $c_M=2D$ (see Appendix), this procedure of covariant quantisation does not provide a condition for the value of $D$. One has to consider other methods of quantization to answer the question. 


\section{The Induced Vacuum}\label{Iv}
Perhaps the most natural question to ask from the point of view of the tensionless limit is what happens to the original tensile theory under a tensionless limit. So in this section, we follow the evolution of states of the tensile Hilbert space under the tensionless limit. In particular, we will focus on the evolution of the tensile vacuum $|0\>_\a$ as we take tension to zero. It will turn out that the evolved vacuum gives rise to states following the induced representations of the BMS$_3$ algebra, where rather remarkable physics emerges.

\subsection{The vacuum from the limit}
Let us now concentrate on the vacuum of the tensile string theory. We can reverse the relations (\ref{ABa}) between the tensile and the tensionless modes:
\be{}
\a_n=\frac{1}{2}\Big[\sqrt{\e}A_n+\frac{1}{\sqrt{\e}}B_{n}\Big],\quad\ta_n=\frac{1}{2}\Big[-\sqrt{\e}A_{-n}+\frac{1}{\sqrt{\e}}B_{-n}\Big] .
\ee
We can put these relations back to write the conditions (\ref{alphvac}) as
\be{limitavac}
\left(\sqrt{\e} A_n+ \frac{1}{\sqrt{\e}} B_n\right)|0\>_\a=0, \ \left(-\sqrt{\e} A_{-n} + \frac{1}{\sqrt{\e}} B_{-n}\right)|0\>_\a=0\quad(n>0).
\ee
We will assume, in keeping with \refb{ms}, that as $\e\to 0$, the tensile vacuum goes to the {\em induced vacuum} $|0, k^\mu\>_I$: 
\be{ai} \lim_{\e\to0} |0, k^\mu\>_\a = |0, k^\mu\>_I.\ee 
This new vacuum is defined as: 
\be{bvac} B_n|0,k^\mu\>_I=0\quad\forall\ n\neq 0, \quad B^\mu_0|0,k^\mu\>_I=k^\mu|0,k^\mu\>_I .\ee
We would also like to mention here that we can also have a special case where $B^\mu_0|0,k^\mu\>_I=0$ which represents the vacuum with zero momentum ($|0, 0^\mu\>_I$). The BMS constraints are very stringent on this vacuum with $L_n|0,0^\mu\>_I=M_n|0,0^\mu\>_I=0\ \ (\forall\ n)$. This corresponds to case (2,2) of table \refb{t1} that we discussed in the previous section. Since there is only one such state that falls in this category, we are going to focus on the case where $k^\mu\neq 0$. Let us now recall the induced BMS representations \refb{ind} 
\bes
\bea{seven.5}
&& M_0 |M,s\> = M |M,s\>, \, L_0 |M,s\> = s |M,s\>; \\ 
&& M_n |M,s\> = 0\quad(n\neq0).
\eea\ees
In terms of oscillator modes these become 
\be{} 
B_0^2 |M, s\> = M |M, s\>;\quad B_n |M, s\> = 0\quad(n\neq0). \ee
Identifying $|M, s\>$ with $|0, k^\mu\>_I$ we observe that this vacuum in the induced representation and hence the name. The physical state conditions \refb{5.89b} on the induced vacuum are partially realised by
\be{Minduced}
M_n|0,k^\mu\>_I=\sum_m B_{-m}\cdot B_{m+n}|0,k^\mu\>_I=0.
\ee
Since $B_n$'s commute amongst each other, there is no ordering ambiguity for the operator $M_0$, i.e. $a_M=0$. So the mass of the vacuum $|0, k^\mu\>_I$ is zero: 
\bea{vacmass}
M_0 |0, k^\mu\>_I &=& \sum_n B_{-n} \cdot B_{n} |0, k^\mu\>_I= \Big(\sum_{n\neq0} B_{-n} \cdot B_{n} + B_0^2 \Big) |0, k^\mu\>_I \nonumber\\
&=& B_0^2 |0, k^\mu\>_I = k^2 |0, k^\mu\>_I =0. \nonumber
\eea
The other constraint condition from \refb{5.89b} for $L_n$ needs to be imposed in its sandwich form:
\be{Linduced}
{}_I\<0,k^\mu|L_n|0,k^\mu\>_I={}_I\<0,k^\mu|A_{n}\cdot B_0|0,k^\mu\>_I= c'k\cdot{}_I\<0,k^\mu|A_{n}|0,k^\mu\>_I=0.
\ee
The above conditions help us to compute the central charges for the induced representation (Appendix \ref{A3}). We obtain their values as
\be{} c_L=c_M=0, \ee
In terms of $C$ oscillators, the induced vacuum conditions have a suggestive form,
\be{seven.6}
(C^{\mu}_n + \C^{\mu}_{\ -n}) |0\>_I = 0\quad(n\neq 0). 
\ee
This is nothing but the condition defining a {\em{Neumann boundary state}} and the solution is given by 
\be{indvac}
|0\>_I = \mathcal{N} \prod_{n=1}^{\infty} \exp \left( - \frac{1}{n} C_{-n} \cdot \C_{-n} \right) |0\>_c.
\ee
where $ \mathcal{N}$ is a normalisation constant. Now let us see how this can be related to the tensile vacuum $\za$.
We remind ourselves the relation between the $C$ oscillators and the $\a$ oscillators can be obtained as an inverse Bogoliubov transformation from \refb{c1}:
\bea{7.8}
\a^{\mu}_n &=& e^{i G} C_{n} e^{-iG} =\cosh\theta \ {C}^{\mu}_n -\sinh\theta \ \tilde{C}^{\mu}_{- n}; \\
\tilde{\a}^{\mu}_n&=& e^{i G} \tilde{C}_{n} e^{-iG}=-\sinh\theta \ {C}^{\mu}_{-n}+\cosh\theta \ \tilde{C}^{\mu}_{ n}, \nonumber
\eea
where the infinitesimal parameter $\theta$ was related in turn to the limiting parameter $\e$ and generator of the transformation takes the form, 
\be{}
G = i \sum_{n=1}^{\infty} \theta \left[ C_{-n}.\tilde{C}_{-n} -C_n .\tilde{C}_n\right], \ \tanh \theta = \frac{\e-1}{\e+1}. 
\ee
We can use this to relate the tensile and tensionless vacua: 
\bea{}
\za &=& \exp[i G] |0\>_c \\
&=& \left(\frac{1}{\cosh\theta}\right)^{1+1+\hdots} \prod_{n=1}^{\infty} \exp\Big[\frac{\tanh\theta}{n} C_{-n}\tilde{C}_{-n}\Big] |0\>_c. \nonumber
\eea
Using the regularisation: $1+1+1+\hdots \infty=\zeta(0)=-\frac{1}{2}$, we finally get
\be{sqz}
|0\>_\a= \sqrt{\cosh\theta} \prod_{n=1}^{\infty} \exp\left[\frac{\tanh\theta}{n} \, C_{-n} \cdot \tilde{C}_{-n}\right] |0\>_c.
\ee
From the point of view of $|0\>_c$, $|0\>_\a$ is a squeezed state. 
 
\subsection{Emergence of open string} 

In the above expression for the tensile closed string vacuum \refb{sqz}, we get $|0\>_\a=|0\>_c$ at $\e=1$. However as $\e$ is dialled to zero (or equivalently $\tanh\theta=-1$), the vacuum can be written as 
\be{}
\lim_{\e\to0}|0\>_\a= {\mathcal{N}} \prod_{n=1}^{\infty} \exp\left[- \, \frac{1}{n} C_{-n} \cdot \tilde{C}_{-n}\right] |0\>_c,
\ee
where we have used $\lim_{\e\to0}\sqrt{\cosh\theta}=\mathcal{N}$. This is exactly the relation \refb{indvac} that we found from the conditions on the induced vacuum. Hence we have shown that the closed tensile string vacuum $|0\>_\a$ evolves into the induced vacuum  $|0\>_I$ in the extreme limit, which turns out to be a Neumann boundary state. This state is interpreted as an open string which is free to move in all spacetime dimensions. A detailed intuitive analysis of this can be found in \cite{Bagchi:2019cay}.

For non zero $\e$ we still have two copies of Virasoro as the symmetry algebra. At $\e=0$ however the symmetry algebra becomes BMS$_3$. We obtained the central charge $c_M=0$, hence there a truncation of the BMS$_3$ algebra to its single Virasoro sub-algebra is possible \cite{Bagchi:2009pe}. Indeed, the symmetry algebra for an open string is a single Virasoro, therefore from an algebraic perspective the appearance of the open string in the tensionless limit is justified. 

\subsection{Comments on the spectrum}
Having understood how to impose constraints in this case, we now move on to some discussions about the spectrum of the tensionless theory around the induced vacuum. 

\paragraph{Worldsheet condensation of perturbative DOF:} In \cite{Bagchi:2019cay}, the spectrum of the tensile bosonic closed string theory was followed in the tensionless limit. It was shown that any perturbative state in the tensile theory in the limit of $\e\to0$ is reduced just to the induced vacuum: 
\be{} 
|\Psi\> = \lim_{\e\to 0} \rho_{\mu\nu} \ \a^\mu_{-n} \ta^\nu_{-n} |0\>_\a = K |0\>_I.
\ee
In the above equation, $|\Psi\>$ is the tensionless version of the tensile closed string perturbative state at level $n$ and $K$ is a level dependent constant. This novel phenomenon was conjectured to be a Bose-Einstein like condensation of closed string modes to form a long open string in the tensionless limit. All perturbative degrees of freedom of the tensile closed string theory thus vanish and there is an emergent open string in this limit. This was also connected to the Hagedorn transition on strings at very high temperatures. We point the reader to \cite{Bagchi:2019cay} for further details of this analysis and the physical picture of how the open string emerges. 

\medskip 

\paragraph{Emergent non-perturbative DOF?}
The condensation of all closed string degrees of freedom on the worldsheet to form a long open string is obviously very appealing. But we want to ask if we can do more. Is it possible to find some new emergent degrees of freedom that arises at this phase transition point? 

\smallskip

With this question in mind, we will go back to the oscillator construction around this vacuum, in terms of the modes $(A, B)$. At first sight, it seems natural to build states with just $A$ oscillators. However, there is a major caveat. Let us remember that $[A_m,A_n]=0$, and hence the norm of a state $\sim A_{-1}|0\>_I$ is not well defined. We could impose $A_n|0\>_I=0$ $\forall\ n$, but that would lead to all $L_n$ being zero and we will be left with only the vacuum. Therefore we can build excited states using the $A_n$'s. Let us study the action of the constraint $M_n$ on a generic state $A^{\mu_1}_{m_1}\hdots A^{\mu_k}_{m_k}|0,k^\mu\rangle_{I}$. 
\be{generic}
M_n\ \prod_j A^{\mu_j}_{m_j}|0\rangle_{I}=\sum_{k} B_{-k}\cdot B_{k+n}\prod_j A^{\mu_j}_{m_j}|0,k^\mu\rangle_{I}
\ee
This is not necessarily annihiliated since for each $n+m_j=0$ we will get a $B_0$ term. Let us illustrate this for a simple case:  
\be{}
M_n\ A^{\mu}_{-n}|0,k^\mu\>_I=-nB^\mu_0|0,k^\mu\>_I\neq 0. 
\ee
This means that not all states built by the $A_n$'s are physical. It can be checked however that if we build states with the $L_n$s then we can have a non trivial state for which the $M_n$ condition holds. For example,
\be{}
M_n\ L_{-n}|0,k^\mu\>_I=-nB^2_0|0,k^\mu\>_I,
\ee
is satisfied for $L_{-n}|0,k^\mu\>_I$ to be a massless state. Similarly we can have physical states of the generic form 
\be{}
L_{n_1}L_{n_2}\hdots L_{n_k}|0,k^\mu\>_I,
\ee
which are massless. 
However, even for these states, the norm is not well defined due to non-trivial action of the $A_n$ oscillators. 

\smallskip

But there could also be non-perturbatively defined states on this vacuum, which can have well defined norm. For example, one might consider applying a unitary operator $U_n(A,B)$ built out of the oscillators on a generic states to create a tower of excitations. Since we want this state to be both  \textit{non zero and physical}, we must have a combination of $A$ and $B$ in this operator. 

\smallskip

Rather intriguingly, it turns out that the kind of states mentioned above can also exist in the spectrum. Returning to the BMS induced representations, these particular states can be built out of $|M,s\>$, defined in \cite{Barnich:2014kra} as
\be{}
|\varphi\> = \exp\left(i~ \sum_n  \w_n L_n\right) |M,s\>
\ee
where $\w^*_n = \w_{-n}$ are complex coefficients. For our purpose, it could be checked that this state satisfies physical states conditions. Clearly this makes these states  eigenstates of supermomenta $M_n$.  
In terms of the induced vacuum of the tensionless string, these states can be explicitly written as 
\be{}
|\varphi\> = \exp\left( i~\sum_{n,m}  \w_n A_{n-m}\cdot B_m \right) |0, k^\mu\>_I.
\ee
It is important to note that the ground state momentum is not zero here, and this ensures the existence of such a unique state. It is clear that these states have the same mass as $|0\>_I$ and  hence are massless ($k^2 = 0$) as well. These are highly \textit{non-perturbative degrees of freedom}, very different in nature from all the perturbative closed string excitations. 
These states definitely warrant more investigation from worldsheet point of view. We aim to return to this in the near future.

 \section{The Flipped Vacuum}\label{flip}

In this section we are going to study the Hilbert space built on a vacuum by considering a ``flip'' between the creation and annihilation operators of one sector of the oscillators similar to the tensile analogue considered in sub-section \refb{F11}. We will see how this allows us to explicitly impose the right-hand side action of the constraints to build a highest weight representation.

 Let us see how we can achieve this from a limit from the parent theory with flipping. As discussed earlier, the two dimensional Virasoro algebra admits an automorphism given by
$
\bar{\mathcal{L}}_n\rightarrow\bar{\mathcal{L}}'_n=-\bar{\mathcal{L}}_{-n}.
$
In the tensile version it meant that the vacuum was asymmetrically defined for right and left sectors according to \refb{flipvac}. Similarly, 
for the case of tensionless strings, with the help of oscillator construction, we can define the \textit{flipped} vacuum in terms of the oscillators $C$ and $\C$ as \footnote{In the $A,~B$ oscillator picture this vacuum can be equivalently described by demanding all positive modes of both $A$ and $B$ annihilate the vacuum, i.e.
\be{}
A_n|0\>_A=B_n|0\>_A=0\quad(n>0). \nonumber \ee
This can be exactly identified with the first case in (\ref{Q2}). We discuss this vacuum in the oscillator construction for clarity. }
\be{f1} C_n|0\>_A=\C_{-n}|0\>_A=0\quad(n>0). \ee
Let us look at the transformations in the limit \refb{c1} which preserves the commutator brackets of the oscillators. Note that due to the flip, the Bogoliubov transformations don't have a mixing of creation and annihilation operators and the vacua \refb{flipvac} and \refb{f1} are identical. It might be convenient to redefine the flipped oscillator sector such that
\be{f2} \tilde{\mathscr{C}}_n=\C_{-n}. \ee
Then the commutation relations of the oscillators are given by
\be{f3} [C^\mu_m,C^\nu_n]=m\delta_{m+n}\eta^\mn,\quad[\cc^\mu_m,\cc^\nu_n]=-m\delta_{m+n}\eta^\mn,\quad[C^\mu_m,\cc^\nu_n]=0. \ee
Notice the negative sign in the second commutator. With this definition, we can use the negative modes of $C$ or $\cc$ to build up the Hilbert space on this vacuum $|0\>_A$. 
The generators $L_n$ and $M_n$ can be written in terms of the redefined oscillators following \refb{c7} as
\begin{subequations}\label{f3a}
\bea{}
L_n&=&\frac{1}{2} \sum_m \big[ C_{-m}\cdot C_{m+n} - \cc_{-m}\cdot\cc_{m-n} \big],\\
M_n&=&\frac{1}{2} \sum_m \big[ C_{-m}\cdot C_{m+n} + \cc_{-m}\cdot\cc_{m-n} + 2C_{-m}\cdot\cc_{-m-n} \big].
\eea \ees
Let us look at the action of the generators on the vacuum $|0\>_A$, which simply boils down to using right hand action of the constraint operators,
\be{f4} L_n|0\>_A=M_n|0\>_A=0\quad(n>0). \ee
This gives us a hint that the physical states fall into the highest weight representation of BMS$_3$.
We must comment here that in the limiting sense, the appearance of the highest-weight representation is not surprising here at all. As we have explained before, this ``flipped'' string theory appears when we take the ultra-relativistic limit on a twisted string theory defined by physical state conditions (\ref{ambicond}). Due to this $\bar{\mathcal{L}}_n\rightarrow-\bar{\mathcal{L}}_{-n}$ symmetry, the limit effectively becomes that of a non-relativistic one,
and highest weight representations of BMS$_3$ algebra emerges naturally.
 
As detailed earlier, the guiding version of the constraints acting on physical state $|phys\>$ should be governed by the sandwich conditions:
\begin{subequations}\label{f5}
\bea{} \<phys'|(L_n-a_L\delta_{n,0})|phys\>&=&0\quad(n\in\mathbb{Z}); \\
\<phys'|(M_n-a_M\delta_{n,0})|phys\>&=&0\quad(n\in\mathbb{Z}), \eea \end{subequations}
where $a_L$ and $a_M$ are ordering ambiguities. Since we are looking for highest weight representations we can restrict the constraint conditions to
\begin{subequations}\label{f6}
\bea{} (L_n-a_L\delta_{n,0})|phys\>&=&0\quad(n\geq0), \\
(M_n-a_M\delta_{n,0})|phys\>&=&0\quad(n\geq0). \eea \end{subequations}
In this case, normal ordered zero modes are defined by
\begin{subequations}\label{f7}
\bea{} L_0&=&\sum_{m>0}\Big[C_{-m}\cdot C_{m}-\cc_{-m}\cdot\cc_{m}\Big]=N+\tilde{N}, \\
M_0&=&2C_0^2+\sum_{m>0}\Big[C_{-m}\cdot C_{m}+\cc_{-m}\cdot\cc_{m}+C_{-m}\cdot\cc_{m}+\cc_{-m}\cdot C_{m}\Big] \nonumber\\
&=&c'k^2+N-\tilde{N}+X+Y,
\eea
\end{subequations}
which appears to be very similar to the oscillator vacuum discussed in chapter \ref{osc}, $N$ counts the number of $C$ modes in the standard way. $\tilde{N}$ has an extra sign in this case due to the sign in the commutator of $\cc$. $X$ and $Y$ are again coupled operators, but unlike in the case of $\zc$ they are not ``bad''. It is indeed possible to obtain eigenvalues for $M_0$ acting on certain states. This is a unique feature of this particular vacuum, as the ``flip'' enables us to efficiently use the machinery of highest weights. We can calculate the values of the central charges of BMS$_3$ for this vacuum (Appendix \ref{A2}) which turn out to be
\be{f7} c_L=2D,\quad c_M=0. \ee
Curiously, this is exactly opposite to what we got for $\zc$ case (\ref{c14}). This also matches with the limit from the tensile case where we have $c=\bar{c}=D$, such that
\be{f7a} c_L=c+\bar{c}=2D\quad and\quad c_M=\e(c-\bar{c})=0. \ee
In what follows, we will discuss the physical states and spectrum built on $|0\>_A$.
 
\subsection{Physical States}

We now need to find the possible physical states allowed by the constraints. Before we start a level by level analysis let us look at the $L_0$ constraint on a physical state in particular.  
\be{}
(L_0-a_L)|phys\>=(N+\tilde{N}-a_L)|phys\>=0. 
\ee
This condition is satisfied if the number operators add up to the ordering ambiguity. This tells us that the physical spectrum is truncated at the particular level that matches the value of $a_L$. This was also seen in the tensile analogue of the flipped vacuum \cite{Lee:2017utr}, where the number of the levels add up to $a+\bar{a}$, the sum of the two ambiguities.  

\paragraph{Level 0:} Let us consider the vacuum $|0\>_A$.
For $n>0$ the conditions \refb{f5} are trivially satisfied, while the zero modes give 
\begin{subequations}\label{f8}
\bea{}
(L_0-a_L)|0\>_A&=&(N+\tilde{N}-a_L)|0\>_A=0, \\
(M_0-a_M)|0\>_A&=&(c'k^2-a_M)|0\>_A=0. \eea
\end{subequations}
We observe that the vacuum will be a physical state if we have $a_L=0$ while the value of $a_M$ will give us the mass. 

\paragraph{Level 1:} If we start with a state $C_{-1}|0\>_A$, then $M_0$ will not be an eigen-valued operator, since the operator $X$ maps $\cc_{-1}|0\>_A$ to $C_{-1}|0\>_A$ while $Y$ does the reverse. Hence we need a linear combination of both states to define a consistent level 1 state: 
\be{f8a} |1\>=a_\mu C^\mu_{-1}|0\>_A +b_\mu\cc^\mu_{-1}|0\>_A. \ee
The $L_0$ condition simply counts the level. The other conditions required for this to be a physical state are
\begin{subequations}\label{f9}
\bea{} L_1|1\>&=&c'k\cdot(a+b)|0\>_A=0, \\
 M_1|1\>&=&2c'k\cdot(a-b)|0\>_A=0, \\
 (M_0-a_M)|1\> &=&\big[(c'k^2-a_M+1)a-b\big]\cdot C_{-1}|0\>_A \nonumber \\
 &+&\big[(c'k^2-a_M-1)b+a\big]\cdot \cc_{-1}|0\>_A=0. \eea
\end{subequations}
Putting $x=(c'k^2-a_M)$, from the last condition we have $(1+x)a_\mu=b_\mu$ and $(1-x)b_\mu=a_\mu$. This is simultaneously possible only when $x=0$, provided $a_\mu=b_\mu\neq0$. The other conditions give us $k\cdot a=0$. Interestingly, the norm of this state can be calculated as $\<1|1\>=a^2-b^2=0$, since we put $a_\mu=b_\mu$. Therefore we find a null physical state at this level.

\paragraph{Level 2:} Here again we need to consider the most general states with all combinations, and in level 2 we can have a six-element basis to generate a state,
\bea{f10} |2\>&=&a_\mu C^\mu_{-2}|0\>_A+e_{\mu\nu} C^\mu_{-1}C^\nu_{-1}|0\>_A+h_{\mu\nu} C^\mu_{-1}\cc^\nu_{-1}|0\>_A \cr
&+&b_\mu\cc^\mu_{-2}|0\>_A+f_{\mu\nu}\cc^\mu_{-1}\cc^\nu_{-1}|0\>_A+j_{\mu\nu} C^\mu_{-1}\cc^\nu_{-1}|0\>_A. 
\eea
Here, $e_{\mu\nu}$ and $f_{\mu\nu}$ are symmetric by construction. For the cross term, we assume $h_{\mu\nu}$ is symmetric while $j_{\mu\nu}$ is anti-symmetric.
Now let us apply the physical state constraints to get the following conditions:
\begin{subequations}\label{f11}
\bea{} L_2|2\>&=&\Big[2c'k\cdot(a+b)+\frac{1}{2}(e^\mu_{\ \mu}-f^\mu_{\ \mu})\Big]|0\>_A=0,  \\
 M_2|2\>&=&\Big[4c'k\cdot (a-b)+(e^\mu_{\ \mu}+f^\mu_{\ \mu})-h^\mu_{\ \mu}\Big]|0\>_A=0, \\
L_1|2\>&=&\Big[2a_\nu +c'(e_{\mu\nu}+h_{\mu\nu}-j_{\mu\nu})k^\mu\Big] C^\nu_{-1}|0\>_A \nonumber \\
&+&\Big[2b_\nu+c'(f_{\mu\nu}+h_{\mu\nu}+j_{\mu\nu}) k^\mu\Big]\cc^\nu_{-1}|0\>_A=0, \\
M_1|2\>&=&2\Big[(a_\nu -b_\nu )+c'(e_{\mu\nu}-h_{\nu\mu}-j_{\mu\nu})k^\mu\Big]C^\nu_{-1}|0\>_A \nonumber \\
&+&2\Big[(a_\nu-b_\nu)-c'(f_{\mu\nu}-h_{\mu\nu}-j_{\nu\mu})k^\mu\Big]\cc^\nu_{-1}|0\>_A=0. 
\eea
\end{subequations}
The $L_0$ condition give us $a_L=2$. Applying the other condition for zero modes we get
\bea{f12}
(M_0-a_M)|2\>&=&\Big[(c'k^2-a_M+2)a_\mu-2b_\mu \Big]C^\mu_{-2}|0\>_A \nonumber \\
&+&\Big[(c'k^2-a_M-2)b_\mu+2a_\mu\Big] \cc^\mu_{-2}|0\>_A \nonumber \\
&+&\Big[(c'k^2-a_M+2)e_{\mu\nu}-h_{\mu\nu}\Big] C^\mu_{-1}C^\nu_{-1}|0\>_A \nonumber \\
&+&\Big[(c'k^2-a_M-2)f_{\mu\nu}+h_{\mu\nu}\Big]\cc^\mu_{-1}\cc^\nu_{-1}|0\>_A \nonumber \\
&+&\Big[(c'k^2-a_M)(h_{\mu\nu}+j_{\mu\nu}) +2e_{\mu\nu}-2f_{\mu\nu} \Big]C^\mu_{-1}\cc^\nu_{-1}|0\>_A=0.
\eea
We can solve for $(c'k^2-a_M)$ in the same way as we did for level 1 and obtain $a_M=c'k^2$, $a_\mu=b_\mu$. Solving \refb{f11} systematically yields the relations $\frac{1}{2}h_{\mu\nu}=e_{\mu\nu}=f_{\mu\nu}$; $e_{\mu\nu}k^\nu=j_{\mu\nu}k^\nu=0$ and $a_\mu=0$. Therefore we can write down the resulting level 2 state as :
\bea{f13} |2\>&=&
e_{\mu\nu}\Big[C^\mu_{-1}C^\nu_{-1}|0\>_A+2C^\mu_{-1}\cc^\nu_{-1}|0\>_A+\cc^\mu_{-1}\cc^\nu_{-1}|0\>_A\Big] \ \ \rightarrow\text{Symmetric} \cr
&+&j_{\mu\nu} C^\mu_{-1}\cc^\nu_{-1}|0\>_A \ \ \rightarrow\text{Anti-symmetric}. 
\eea
Similar to the level 1 analysis, we find that the norm of this state also vanishes. Looking at the spectrum, as we have mentioned, there are three distinct objects, a scalar (trace part from $e_{\mu\nu}$), a symmetric traceless and an anti-symmetric tensor state (also see \cite{Gamboa:1989px,Casali:2016atr}). It is almost tempting to compare these states with the tensile string physical state spectrum, but without proper understanding of the symmetries of background spacetime, we refrain from doing so.


\subsection{Null States}

Let us consider GCA$_2$/BMS$_3$ highest weight null states and make a connection to those discussed in \cite{Bagchi:2009pe}. At level 1 we can write a general null state,
\be{} |1\>=\gamma_1 L_{-1}|0\>_A+\gamma_2 M_{-1}|0\>_A, \ee
where $\gamma$'s are numerical constants. For $n>1$ the physical conditions are trivially satisfied for this state. This is a physical null state if $L_1|1\>=M_1|1\>=0$ which translates to the following, 
\be{} \gamma_1=0\quad and \quad a_M=0. \ee
In terms of oscillators this level 1 state is given as,
\be{} |1\>=M_{-1}|0\>_A=2k\cdot (C_{-1}+\cc_{-1})|0\>_A, \ee
which we have also obtained in \refb{f8a} with the identification $a_\mu=k_\mu$. At level 2 we can write a general null state as 
\bea{}
|2\>&=&\big[\gamma_1L_{-2}+\gamma_2M_{-2}+\gamma_3L_{-1}L_{-1}+\gamma_4M_{-1}M_{-1}+\gamma_5L_{-1}M_{-1}\big]|0\>_A .
\eea
If we require that $|2\>$ is a physical state then this should be annihilated by modes of $L_n$ and $M_n$ up to $n=2$ (beyond $n>2$ it is trivially $0$). The results that we would obtain would be identical to the discussion in  \cite{Bagchi:2009pe} with $\Delta=a_L$, $\xi=a_M$, $C_1=\frac{D}{6}$ and $C_2=0$. We spell out these conditions here for convenience:
\begin{subequations}\label{}
\bea{}
3\gamma_1+2(2a_L+1)\gamma_3+2a_M\gamma_5&=&0, \\
(4a_L+D)\gamma_1+6a_L\gamma_3+6a_M\gamma_5+4a_M\gamma_2&=&0, \\
2(a_L+1)\gamma_5+4a_M\gamma_4+3\gamma_2&=&0, \\ 
4a_M\gamma_1=a_M\gamma_3&=&0, \\
3\gamma_1+2\gamma_3+2a_M\gamma_5&=&0.
\eea
\end{subequations}
Solving the set of equations and following the analysis of \cite{Bagchi:2009pe}, we will find a non trivial state exists only for $a_M=0$, $\gamma_1=\gamma_3=0$, $\gamma_2=-\frac{2a_L+1}{3}\gamma_5$, provided $a_L\neq -\frac{1}{4}D$. If we break down $|2\>$ in terms of oscillators and apply the conditions of \refb{f11}, we will also get an additional constraint $\gamma_2=-\gamma_4$. Finally, collecting all the parameters, we can write this null state as a whole,
\bea{}
|2\>&=&M_{-2}|0\>_A-M_{-1}M_{-1}|0\>_A-\frac{3}{2a_L+1}L_{-1}M_{-1}|0\>_A 
\eea

This is clearly a linear combination of 2D GCA null states as derived intrinsically in \cite{Bagchi:2009pe}. Moreover, one can see that these states appear as the UR limit of the null states in the parent flipped theory as discussed in section \refb{F11}. The limiting null states will be given by,
  \be{}
  |\chi_1\rangle = \lim_{\e\to 0}~  |\chi_R\rangle-  |\chi_L\rangle,~~~ |\chi_2\rangle = \lim_{\e\to 0}~ \e( |\chi_R\rangle+ |\chi_L\rangle).
  \ee
  Taking the limits consistently will lead us to,
  \bea{}
   |\chi_1\rangle &&\sim-  M_{-1}M_{-1}|\phi\rangle\nonumber \\
    |\chi_2\rangle &&\sim \left( M_{-2} - \eta L_{-1}M_{-1}\right) |\phi\rangle .
  \eea
Which can be checked to be 2D GCA null states mentioned in  \cite{Bagchi:2009pe} for the case $h_M=0$ in the NR limit.  So all in all we take an UR limit on the states, but arrive at the well-known NR answer as in  \cite{Bagchi:2009pe}, keeping with the spirit of the flipped theory. 

We must note here that this analysis doesn't give us a fixed level since $a_L$ is undetermined. However, if we take the limit from the tensile case, we have the two vacua to be identical to each other. This enables us to map the ambiguities and the central charges directly. We find that $a_L=a+\bar{a}=2$ while $a_M=\e(a-\bar{a})=0$ with respect to the Virasoro ordering ambiguities. The central charges follow in same way in the limit \refb{f7a}. This means we see a tensionless  theory in $D=26$ dimensions with limited massless spectrum at level 2, in a similar way as in \cite{Huang:2016bdd}.

\subsection{Taking limit from parent states}

A supposed problem with the spectrum we described for the flipped tensionless strings is all the physical states seem to be null in this case. This seems to create a confusion since the parent states had definite norms as discussed in section (\ref{F11}), albeit with some added peculiarities.  One must understand how these states arise in the limit in order to say more about the spectrum. 

In this section we would concentrate on the parent massless state $|\psi_1\rangle= \xi_{\mu\nu}\alpha_{-1}^{\mu}\overline{\alpha}_{+1}^{\nu}|0\rangle_A$ which has a positive norm coming from the symmetric part of $\xi_{\mu\nu}$ as described before in (\ref{F11}).  Using the limiting relations between $\a$ and $C$ oscillators \refb{c1}, we can write the fate of the state in the limit, 
\bes\bea{}
\lim_{\eps\to0}|\psi_1\rangle&=&\xi_{\mu\nu}\big[\cosh \theta \ C^\mu_{-1}-\sinh \theta \ \cc^\mu_{-1}\big]\big[\sinh \theta \ C^\nu_{-1}-\cosh \theta \ \cc^\nu_{-1}\big]|0\rangle_A. \\
&=&\gamma_1|\phi\>_1+\gamma_2|\phi\>_2+\gamma_3|\phi\>_3,
\eea\ees
where the different excitations are,
\bes\bea{}
|\phi\>_1&=&\e\xi_{\mu\nu}\Big[C^\mu_{-1}C^\nu_{-1}-2C^\mu_{-1}\cc^\nu_{-1}+\cc^\mu_{-1}\cc^\nu_{-1}\Big]|0\rangle_A \nonumber\\
|\phi\>_2&=&\frac{1}{\e}\xi_{\mu\nu}\Big[C^\mu_{-1}C^\nu_{-1}+2C^\mu_{-1}\cc^\nu_{-1}+\cc^\mu_{-1}\cc^\nu_{-1}\Big]|0\rangle_A \nonumber\\
|\phi\>_3&=&\xi_{\mu\nu}\Big[C^\mu_{-1}\cc^\nu_{-1}-C^\nu_{-1}\cc^\mu_{-1}\Big]|0\rangle_A,
\eea\ees
which evidently occur at different orders of $\e$. For any arbitrary $\e$, the coefficients $\gamma_1$, $\gamma_2$ and $\gamma_3$ can be normalised to be $\frac{1}{2\sqrt{2}}$, $\frac{1}{2\sqrt{2}}$ and $1$. Comparing this to the level two physical state \refb{f13} we find the symmetric and the anti symmetric parts at order $\e^{-1}$ and $\e^0$ respectively by redefining $e_{\mu\nu}=\frac{1}{\e}\xi_{\mu\nu}$ and $j_\mn$ as the anti-symmetric part of $\xi_\mn$. It should be noted that anti-symmetric part of the parent state had zero norm, and in the limit it exactly reduces to another antisymmetric state ($|\phi\>_3$) with zero norm.

 Actually, one can check that the norm of each state $|\phi\>_1$, 
$|\phi\>_2$ or $|\phi\>_3$ are individually zero owing to the sign in the commutator of $\cc_n$ oscillators. However the norm of the total state $|\psi\>_1$ in the limit is non-zero and can be calculated by the following, where we only list the contributing terms coming from the inner product of $\<\phi_1|\phi_2\>$,
\bea{}
\<\psi_1|\psi_1\>&=&\frac{1}{8}\xi_\mn\xi_{\rho\s} {}_A\<0|\Big[C^\mu_{1}C^\nu_{1}C^\rho_{-1}C^\s_{-1}-4C^\mu_{1}\cc^\nu_{1}C^\rho_{-1}\cc^\s_{-1}+\cc^\mu_{1}\cc^\nu_{1}\cc^\rho_{-1}\cc^\s_{-1}\Big]|0\>_A \cr
&=&\frac{1}{8}\xi_\mn\xi_{\rho\s} {}_A\<0|\Big[2\eta^{\mu\rho}\eta^{\nu\s}+4\eta^{\mu\rho}\eta^{\nu\s}+2\eta^{\mu\rho}\eta^{\nu\s}\Big]|0\>_A =\xi_\mn\xi^\mn.
\eea

So in general the norm of the total state remains preserved under the limit. One can see the $\mathcal{O}(\e)$ contribution, i.e. $|\phi\>_1$ is not a physical state combination and hence does not appear in  \refb{f13}. This apparent loss of information made it seem that generic physical state in level two will be null. 

\newpage

\section{Conclusions}
In this paper, we have investigated, in detail, the quantum structure of the tensionless bosonic closed string within the premise of canonical quantisation. We have seen that a careful analysis of the imposition of constraints leads us to three different quantum mechanical theories starting from the same classical null string. We have analysed the different vacua of these theories and some properties of these different theories arising of these diverse vacuum structure. 

\medskip

\noindent There are innumerable directions of future work, some of the most pressing of which we elaborate on below. 
\begin{itemize}

\item{\em All quantum theories:} We have worked under the ambit of canonical quantisation. It would be of importance to see if other types of quantisation also lead us to these three distinct theories, or perhaps one or more of them are ruled out. We want to address lightcone quantisation first and this is being currently pursued. The path integral quantisation poses interesting problems as we need to consider null surfaces and the underlying Riemannian structure of usual string geometry would change in this case. We are hopeful that BRST quantisation would be more accessible and plan to look at this in the near future. 
\end{itemize}

\noindent For each of the different vacua addressed here, there are a number of immediate questions. 

\begin{itemize}

\item {\em The Oscillator vacuum:} We seem to have a huge number of null states in the spectrum around the oscillator vacuum. Especially, all the massless sector that we have described also happens to be null. This is indication of a huge (higher spin) gauge symmetry in the quantum theory. We wish to concretise this and hopefully find links to the recent discourse of tensionless strings and higher spin theory \cite{Sagnotti:2003qa, Bonelli:2003kh}. 

We have also not been able to fix the critical dimensions in our way of formulating the theory. Light-cone quantisation here would come in handy here. 

Another very important question is about representation theory. While the other two vacua are directly linked to known representations of the underlying BMS algebra, the oscillator vacuum does not seem to clearly follow from BMS representation theory. To elaborate this point, note that the Induced vacuum is of course the vacuum of the induced representation, and the flipped vacuum is the vacuum of the highest weight representations. The oscillator vacuum thus could lead us to a hitherto unknown representation of the BMS group. Even if it does not, then it would be very interesting to understand the interplay between the different representations that give rise to the oscillator vacuum. 

\item{\em The Induced vacuum:} The induced vacuum leads to some very intriguing physics, as has been discussed in \cite{Bagchi:2019cay} and we have slightly elaborated this in the present paper. The physics of the Bose Einstein like condensation of states and also the implications of Rindler-like physics on the string worldsheet need a lot of further investigation. One of the primary questions is the formulation of the phase transition near the tensionless point. We would like to characterise this in terms of an order parameter and work out the statistical mechanical picture in more detail. 

It also seems likely that the Gross-Mende like relations of the scattering amplitudes in the very high energy regime of string theory should arise from the quantum mechanical tensionless theory around the induced vacuum. It should be possible to construct vertex operators in a BMS-invariant way and then use them to build up $n$-point amplitudes of states. This can then be compared to answers arising from the tensionless limit of the tensile string scattering amplitudes themselves. 

\item{\em The Flipped vacuum:} The spectrum around the flipped vacuum is also rather weird. Firstly, the spectrum is truncated and the value of $a_L$ and $a_M$ are not fixed in the canonical analysis. The limit suggests an answer and this ties up with what has been found earlier in \cite{Gamboa:1989px, Casali:2016atr}. An independent analysis of this would be appropriate and is currently underway from the point of view of a light-cone analysis. Secondly, even with the truncated spectrum, the physical state conditions seem to conspire in a way so as to make all the existing states null. This is not particularly pleasing. Perhaps this is an indication that the bosonic ambitwistor theory does not make sense, as has been put forward in \cite{Casali:2016atr}. The natural direction would be to move to a supersymmetric theory. 

\end{itemize}

\noindent We also wish to investigate some natural generalisations and other directions. 

\begin{itemize}

\item{\em Superstrings:} Possibly the most natural generalisation of our analysis in the current paper is to look at the supersymmetric version of the current analysis. The process of classifying all possible quantum theories would be somewhat more involved due to the presence of more generators in the underlying SUSY algebra. In this context, it is also worth mentioning that there are two possible choices of the Super BMS (SBMS) algebra that can arise on the worldsheet of the tensionless closed superstring, viz. the homogeneous \cite{Bagchi:2016yyf} and the inhomogeneous SBMS \cite{ Bagchi:2017cte}. Hence, an exhaustive analysis would be even more tedious. But we expect to see some very interesting physics emerging, e.g. in the equivalent of the induced vacuum for the tensionless superstring. It is expected that this will lead to something that is qualitatively different from the picture of the Bose-Einstein like condensation we found for the bosonic tensionless string.  

\item{\em Transitions between different vacua?} Another immediate question is whether there can be transitions between these various vacua e.g. via instantons. If these are present, these would be novel quantum phase transitions between different tensionless string theories, unlike anything we have encountered before. 

\end{itemize}

These are only some of the directions we wish to make progress on in the near future. It is clear that a very fertile and exciting corner of string theory has been reinvigorated by our recent investigations. We wish to pursue this with enthusiasm and hope to unearth more exciting physics in this extreme limit of string theory. 

\section*{Acknowledgements}
We would like to thank Hamid Afshar, Andrea Campoleoni, Daniel Grumiller, Alejandro Rosabal, Bo Sundborg, Chi Zhang for inspiring discussions. AB is partially supported by the following grants from the Science and Engineering Research Board, India: EMR/2016/008037, ERC/2017/000873, MTR/2017/000740. ArB was supported in part by Chinese Academy of Sciences (CAS) Hundred-Talent Program, Key Research Program of Frontier Sciences, CAS,  Project 11647601 supported by NSFC, The Korea Ministry of Education, Science and Technology, ICT and Future Planning, Gyeongsangbuk-do and Pohang City. SC is partially supported by the ISIRD grant 9-252/2016/IITRPR/708. Pulastya Parekh (PP) was funded by the Junior Research Fellowship Programme from Erwin Schr{\"{o}}dinger International Institute for Mathematics and Physics, Vienna. ArB would like to thank the ICTP HECAP, CERN PH-TH, NCTS Hsinchu and OIST Okinawa for warm hospitality during the course of this project. PP thanks Institute for Theoretical Physics, TU Wien and Centro de Estudios Científicos, Valdivia where a part of this project was completed.

\newpage

\appendix

\section*{Appendices}

\section{Other constraint conditions}\label{ApB}

In this appendix we are going to look at the remaining five constraint conditions possible to impose the parent condition of \refb{Constraint} as discussed in Section \ref{five}. Then we shall show how inconsistencies arise from these conditions and hence the ones considered in the above sections are the only possible way to get a non-trivial and meaningful spectrum of the tensionless string. We are going to show this by considering a vacuum which we assume to obey the physical conditions for at least $n\neq 0$. Referring to table \refb{t1} we consider the following cases: 

\paragraph{Case (1,2):} This condition is given by 
\be{} L_{n}|phys\>=0,\quad\forall\ n>0,\quad M_{n}|phys\>=0,\quad\forall\ n\neq0. \ee
For starting, let us consider a physical vacuum $|0,k^\mu\>$ that satisfies  
\be{}
  M_{n}|0,k^\mu\>=  \sum_m    B_{-m}\cdot B_{m+n}|0,k^\mu\>=0 \quad\forall\ n\neq 0. 
\ee
This immediately tells us that we must have $B^\mu_n|0,k^\mu\>=0$ $(n\neq 0)$. For the other condition we have 
\bea{}
  L_{n}|0,k^\mu\>&=&  \sum_m    A_{-m}\cdot B_{m+n}|0,k^\mu\>=0  \quad\forall\ n>0 \cr
 &=&  A_n\cdot B_0|0,k^\mu\>=0  \quad\forall\ n>0. 
\eea
For the vacuum with generic momentum this gives rise to the additional property $A^\mu_n|0,k^\mu\>=0$ $(n> 0)$. This means the excited states can be defined by 
\be{}
A^{\mu_1}_{-n_1}A^{\mu_2}_{-n_2}\hdots A^{\mu_j}_{-n_i}|0,k^\mu\>,\quad \forall\ n_1,n_2,\hdots, n_i>0.
\ee
Therefore the vacuum is defined by 
\bes\bea{}
A^\mu_n|0,k^\mu\>&=&(C^\mu_n-\C^\mu_{-n})|0,k^\mu\>=0 \quad\forall\ n>0, \\
B^\mu_n|0,k^\mu\>&=&(C^\mu_n+\C^\mu_{-n})|0,k^\mu\>=0 \quad\forall\ n\neq 0,
\eea\ees
where we have used \refb{CC} relating $A,B$ and $C,\C$ oscillators. We can rewrite the above conditions as 
\bes \bea{}
C^\mu_n|0,k^\mu\>&=&\C^\mu_{-n}|0,k^\mu\>=0 \quad\forall\ n>0, \\
C^\mu_n|0,k^\mu\>&=&-\C^\mu_{-n}|0,k^\mu\> \quad\forall\ n<0.
\eea \ees
The first equation is actually the condition for the flipped vacuum $|0\>_A$ defined in \refb{flipvac}. The second condition is an extra restriction that couples the creation modes of $C$ oscillators with creation modes of $\C$. Basically this reduces the system to only one set of oscillators and the realisation of the algebra is somewhat destroyed, since the decoupling of the modes is crucial for the BMS Algebra to close.  

\paragraph{Case (1,3): } This case is represented by 
\be{b3}
 L_{n}|phys\>= 0,\quad\forall\ n>0,\quad M_{n}|phys\>\neq0,\quad\forall\ n.
 \ee
but $\<phys'|M_{n}|phys\>=0$ $\forall\ n$. Here, the only meaningful way we can realise this is choosing $B_n\sim C_n\pm\C_{-n}$ for the oscillator vacuum $\zc$. Once again we find that closure of the BMS Algebra requires $A_n\sim C_n\mp\C_{-n}$. The first condition is realised on the vacuum by $A_n|0,k^\mu\>=0$, $n>0$. This imposes an extra restriction on $\zc$ in the following way
\be{} A_n|0,k^\mu\>_c=0\quad\forall\ n>0\implies C_n|0,k^\mu\>_c=\C_{-n}|0,k^\mu\>_c=0 \quad\forall\ n> 0. \ee
This brings us to case (1,2) in a way, that one of the oscillators fall off, but this time for $\zc$. Once again, the BMS Algebra would break down and hence we end up with an inconsistency. 

\paragraph{Case (2,1):} This case is given by the conditions 
\be{} L_{n}|phys\>=0,\quad\forall\ n\neq 0,\quad M_{n}|phys\>=0,\quad\forall\ n>0. \ee
The first condition can be realised on the physical vacuum by
\be{}
  M_{n}|0,k^\mu\>=  \sum_m    B_{-m}\cdot B_{m+n}|0,k^\mu\>=0 \quad\forall\ n>0, 
\ee
which gives us $B_n|0,k^\mu\>=0$ $(n>0)$. Expanding $L_n$  and imposing the second condition 
\be{}
  L_{n}|0,k^\mu\>=  \sum_m    A_{-m}\cdot B_{m+n}|0,k^\mu\>=0 \quad\forall\ n>0, 
\ee
gives $A_n|0,k^\mu\>$ $(n\neq0)$. Following the same steps as done for case (1,2) we will end up getting the conditions 
\bes \bea{}
C^\mu_n|0,k^\mu\>&=&\C^\mu_{-n}|0,k^\mu\>=0 \quad\forall\ n>0, \\
C^\mu_n|0,k^\mu\>&=&\C^\mu_{-n}|0,k^\mu\> \quad\forall\ n<0.
\eea \ees
We see once again that here the $C$ and $\C$ oscillators in the case of the flipped vacuum become coupled to each other and we are left with only one set of oscillators, and the BMS algebra doesn't close.

\paragraph{Case (2,2):} Now let us consider 
\be{b2}
 L_{n}|phys\>= 0,\quad M_{n}|phys\>=0,\quad\forall\ n\neq0.
 \ee
For a vacuum $|0,k^\mu\>$ to satisfy this, it must obey the following condition.
\be{}
 M_{n}|0,k^\mu\>= \frac{1}{2}\sum_m B_{-m}\cdot B_{m+n}|0,k^\mu\> =0\quad\forall\ n\neq 0.
 \ee
Since all $B_n$s commute, this boils down to $B_n|0,k^\mu\>=0$ $\forall\ n$, which is precisely the condition \refb{bvac}. It immediately follows that the for the condition $L_{n}|phys\>=0$, $\forall\ n\neq0$  to be true we also require $A_n|0,k^\mu\>=0$, $\forall\ n\neq0$. This gives us a case with only the vacuum with no other excitations. Alternatively this can also be achieved by setting $k^\mu=0$ for all the states which combined with \refb{seven.6} is the condition for boundary states. Therefore we see that the cases from $M_{n}|phys\>= 0$, $\forall\ n\neq0$, follow as special cases of the \textit{induced} vacuum discussed extensively in Section \ref{ind}.  

\paragraph{Case (2,3):} Let us consider the following case for imposing the constraint:
\be{b1}
 L_{n}|phys\> = 0,\quad M_{n}|phys\>\neq 0,\quad\forall\ n\neq 0, 
 \ee
but  $\<phys'|M_{n}|phys\>=0$ $\forall\ n$. For starting, let us consider a physical vacuum $|0,k^\mu\>$ that satisfies  
\be{}
  L_{n}|0,k^\mu\>=  \sum_m    A_{-m}\cdot B_{m+n}|0,k^\mu\> \quad\forall\ n\neq 0. 
\ee
which leads to either $A_n|0,k^\mu\>=0$ or $B_n|0,k^\mu\>=0$, $\forall\ n\neq 0$. The latter gives us $M_{n}|0,k^\mu\>=0$ which contradicts \refb{b1}. If we choose the former then the only meaningful way we can realise this is choosing $B_n\sim C_n\pm\C_{-n}$ and the vacuum conditions being \refb{c3}, in order to satisfy the sandwich condition of $M_n$. Then the algebra \refb{Q4} which is vital for the closure of the BMS Algebra dictates that $A_n\sim C_n\mp\C_{-n}$. It can be easily seen that
\be{} A_n|0,k^\mu\>=0\quad\forall\ n\neq 0\implies B_n|0,k^\mu\>=0 \quad\forall\ n\neq 0. \ee
Thus we see that since the oscillators and constraints are intertwined by the commutations that preserve the algebra, we get back from the above  $M_{n}|0,k^\mu\>=0$ which again contradicts \refb{b1}. This tells us that the condition $M_{n}|phys\> \neq 0$, but $\<phys'|M_{n}|phys\>=0$ $(\forall\ n)$ is consistent only with the \textit{oscillator} vacuum defined by $\refb{c3}$.

\paragraph{Case (3,1):} Last, but not the least we turn our attention to the condition 
\be{b}
 L_{n}|phys\>\neq 0,\quad\forall\ n,\quad M_{n}|phys\>=0,\quad\forall\ n>0,
 \ee
but $\<phys'|L_{n}|phys\>=0$ $\forall\ n$. Here again, realising the second condition on the vacuum gives us $B_n|0,k^\mu\>=0$ for $n>0$. The non trivial way to impose the first condition would be 
\be{} \<0,k^\mu|L_n|0,k^\mu\>=\sum_m \<0,k^\mu|A_{-m}\cdot B_{m+n}|0,k^\mu\>=0. \ee
This gives us $\<0,k^\mu|A_n|0,k^\mu\>=0$. We can realise this by choosing $A_n\sim C_n\mp\C_{-n}$ for the oscillator vacuum $\zc$. The logic from case (1,3) would follow and we would end up getting the extra condition on $\zc$ as 
\be{} B_n|0,k^\mu\>_c=0\quad\forall\ n>0\implies C_n|0,k^\mu\>_c=-\C_n|0,k^\mu\>_c=0 \quad\forall\ n> 0, \ee
which again is an inconsistency. \\
\newline
Thus we see that since the oscillators and constraints are intertwined by the commutations that preserve the algebra, the only interesting and self-consistent cases are the \textit{oscillator}, \textit{induced} and \textit{flipped} vacua discussed in the main text above.


\section{Central Charges: Explicit calculations } 
\label{ApA}

The classical Virasoro $\times$ Virasoro algebra satisfied by the constraints $\ML_m$ and $\bL_m$ is
\bea{}
[\ML_m, \ML_n] = (m-n) \ML_{m+n} , \quad [\bL_m, \bL_n] = (m-n) \bL_{m+n}, \quad [\ML_m, \bL_n] = 0.
\eea
When we quantize the system, a central term needs to be added to the classical counterpart of the Virasoro algebra only when we get $\ML_0$ and $\bL_0$ on the right hand side, which we assume to be of the form  
\be{} 
[\ML_m,\ML_{-m}]=2m\ML_0+A(m),~~~
[\bL_m,\bL_{-m}]=2m\bL_0+\bar{A}(m),
\ee
where it can be easily seen that $A(-m)=-A(m)$.  Imposing that these generators satisfy the Jacobi identities, we can derive 
\be{central1}
A(m)=\frac{m(m^2-1)}{6}A(2)-f(m)A(1),
\end{equation}
where we have assumed $f(m)$ to be some function of $m$. A general form of $A(m)$ can be found to be $A(m)=c_1m+c_3m^3$. $\bar{A}(m)$ can be shown to be also of the same form, which enables us to write the algebra with central charges \refb{VIRC}
\bea{ccharge}
[\ML_m, \ML_n] &=& (m-n) \ML_{m+n} + \frac{c}{12} m(m^2-1)\delta_{m+n,0}, \cr
[\bL_m, \bL_n] &=& (m-n) \bL_{m+n} + \frac{\bar{c}}{12} m(m^2-1)\delta_{m+n,0}, \cr
[\ML_m, \bL_n] &=& 0.
\eea
We can fix the central charge $c$ by calculating the value of $A(1)$ and $A(2)$. To do this, we consider the expectation values of the following commutators for the vacuum with zero momentum $|0,0^\mu\rangle$:
\be{} \langle0,0^\mu|[\ML_1,\ML_{-1}]|0,0^\mu\rangle=2\<0,0^\mu|\ML_0|0,0^\mu\>+A(1), \nonumber\ee
\be{AV1} \langle0,0^\mu|\ML_2,\ML_{-2}]|0,0^\mu\rangle=4\<0,0^\mu|\ML_0|0,0^\mu\>+A(2), \ee
with similar expressions for computing $\bar{c}$. This is the basic formulation to calculate the central charges. For the case of BMS, we see a similar structure. A central extension can be constructed for the classical BMS \refb{BMS3} as
\bea{}
[L_m,L_{-m}]&=&2mL_0+A_L(m), \cr
[L_m,M_{-m}]&=&2mM_0+A_M(m).
\eea
The Jacobi identities can be applied here and we find $A_L(m)$ and $A_M(m)$ to have the exact form as \refb{central1}, that are related to the central charges $c_L$ and $c_M$. For $A_L(m)$ this can also be calculated by finding out the following expectation value for the vacuum with zero momentum: 
\be{} \langle0,0^\mu|[L_1,L_{-1}]|0,0^\mu\rangle=2\<0,0^\mu|L_0|0,0^\mu\>+A_L(1) \nonumber\ee
\be{AA1} \langle0,0^\mu|[L_2,L_{-2}]|0,0^\mu\rangle=4\<0,0^\mu|L_0|0,0^\mu\>+A_L(2) \ee
Similarly to find $A_M(m)$, we need to use expectation values  of  $[L_1,M_{-1}]$ and $[L_2,M_{-2}]$.  With some algebra one finds 
\be{} \langle0,0^\mu|[L_1,M_{-1}]|0,0^\mu\rangle=2\<0,0^\mu|M_0|0,0^\mu\>+A_M(1) \nonumber\ee
\be{AA2} \langle0,0^\mu|[L_2,M_{-2}]|0,0^\mu\rangle=4\<0,0^\mu|M_0|0,0^\mu\>+A_M(2) \ee
We are now going to apply the appropriate sandwich conditions for the tensile, oscillator, induced and flipped vacuum to explicitly obtain the central charges. 

\subsection{The tensile vanilla vacuum}\label{ApA0}

Let us consider the tensile case first. The states are governed by two copies of the Virasoro algebra. The constraints act on the vacuum with zero momentum $|0,0^\mu\rangle_\a$ in the following way.\footnote{It is to be noted that the vacuum with zero momentum does not obey the physical state conditions in general.}
\bes\bea{} \ML_0|0,0^\mu\rangle_\a&=&0, \quad \ML_n|0,0^\mu\rangle_\a=0\quad\forall\ n>0, \\
\ML_{-1}|0,0^\mu\rangle_\a&=&\sqrt{\frac{\a'}{2}}k\cdot\a_{-1}|0,0^\mu\rangle_\a=0, \\
\ML_{-2}|0,0^\mu\rangle_\a&=&\frac{1}{2}\a_{-1}\cdot\a_{-1}|0,0^\mu\rangle_\a. 
\eea\ees
Applying these to the conditions \refb{AV1} we obtain 
\be{} {}_\a\langle0,0^\mu|[\ML_1,\ML_{-1}]|0,0^\mu\rangle_\a=0+A(1)=0, \ee
and for $A(2)$, we get  
\bea{ap11} 
{}_\a\langle0,0^\mu|[\ML_2,\ML_{-2}]|0,0^\mu\rangle_\a &=&{} {}_\a\langle0,0^\mu|\ML_2\ML_{-2}|0,0^\mu\rangle_\a=A(2) \nonumber \\
&=& \frac{1}{4} {} {}_\a\langle0,0^\mu|  \a_1\cdot \a_1 \a_{-1}\cdot{\a_{-1}}|0,0^\mu\rangle_\a =\frac{D}{2}.    
 \eea
Collecting the above information for $A(1)$ and $A(2)$ and plugging them back into \refb{central1} we get 
\begin{equation}
  A(m) = \frac{1}{12}D (m^3 - m)
\end{equation}
Likewise by following the same analysis we can show that $\bar{A}(m) = \frac{1}{12}D (m^3 - m)$. Hence from \refb{ccharge} we read off the two central charges as \be{} c=\bar{c}=D. \ee

\subsection{Tensile flipped vacuum} 
As we have mentioned before in section \ref{three} for a flip in the oscillator $\ta_{n}\rightarrow\ta_{-n}$ we will have 
\bes\bea{} \bL_{-n}|0,0^\mu\rangle_A&=&0\quad\forall\ n>0, \\
\bL_{1}|0,0^\mu\rangle_A&=&\sqrt{\frac{\a'}{2}}k\cdot\ta_{1}|0,0^\mu\rangle_A=0, \\
\bL_{2}|0,0^\mu\rangle_A&=&\frac{1}{2}\ta_{1}\cdot\ta_{1}|0,0^\mu\rangle_A. 
\eea\ees
This will modify \refb{ap11} as 
\bea{} 
{}_\a\langle0,0^\mu|[\bL_2,\bL_{-2}]|0,0^\mu\rangle_A &=&- {}_A\langle0,0^\mu|\bL_{-2}\bL_{2}|0,0^\mu\rangle_A=\bar{A}(2) \nonumber \\
&=& -\frac{1}{4} {}_A\langle0,0^\mu|  \ta_{-1}\cdot \ta_{-1} \ta_{1}\cdot{\ta_{1}}|0,0^\mu\rangle_A =-\frac{D}{2}.    
 \eea
This justifies the transformation in the central charge $\bar{c}\rightarrow-\bar{c}$ for the flip. 

\subsection{Tensionless oscillator vacuum}\label{ApA1}
Let us consider $|0,0^\mu\rangle_c$, the oscillator vacuum with zero momentum. The action of $L_0$ and $M_0$ on this state is given from \refb{c20} 
\be{} {}_c\langle0,0^\mu|L_0|0,0^\mu\rangle_c={}_c\langle0,0^\mu|M_0|0,0^\mu\rangle_c=0. \ee
Lets us also remind ourselves of the redefinition we made in \refb{c8}
\bes\bea{}
L_n&=&\J_n-\bJ_{-n} \\
M_n&=&\J_n+\bJ_{-n}+2\K_{n} 
\eea\ees
with $\J_n=\frac{1}{2} \sum_m C_{-m}\cdot C_{m+n}$, $\bJ_n=\frac{1}{2} \sum_m \C_{-m}\cdot\C_{m+n}$ and $\K_n=\frac{1}{2} \sum_m C_{-m}\cdot\C_{-m-n}$. Due to the presence of cross terms in $K_n$ it can be checked that ${}_c\<0,0^\mu|J_m K_n|0,0^\mu\>_c=0$ $\forall$ $m,n$. We proceed to calculate the following commutators for $m=1,2$:
\bea{}
{}_c\langle0,0^\mu|[L_m,L_{-m}]|0,0^\mu\rangle_c&=&{}_c\langle0,0^\mu|L_mL_{-m}|0,0^\mu\rangle_c-{}_c\langle0,0^\mu|L_{-m}L_m|0,0^\mu\rangle_c \cr
&=&{}_c\langle0,0^\mu|\J_{m}\J_{-m}|0,0^\mu\rangle_c-{}_c\langle0,0^\mu|\bJ_{m}\bJ_{-m}|0,0^\mu\rangle_c \cr
&=&A_L(m)=0.
\eea
Similarly we can calculate the commutators for \refb{AA2}
\bea{}
{}_c\langle0,0^\mu|[L_1,M_{-1}]|0,0^\mu\rangle_c&=&{}_c\langle0,0^\mu|L_1M_{-1}|0,0^\mu\rangle_c-{}_c\langle0,0^\mu|M_{-1}L_1|0,0^\mu\rangle_c \cr
&=&{}_c\langle0,0^\mu|\J_{1}\J_{-1}|0,0^\mu\rangle_c+{}_c\langle0,0^\mu|\bJ_{1}\bJ_{-1}|0,0^\mu\rangle_c \cr
=A_M(1)&=&2{}_c\langle0,0^\mu|c'k^2|0,0^\mu\rangle_c=0.
\eea
\bea{}
{}_c\langle0,0^\mu|[L_2,M_{-2}]|0,0^\mu\rangle_c&=&{}_c\langle0,0^\mu|L_2M_{-2}|0,0^\mu\rangle_c-{}_c\langle0,0^\mu|M_{-2}L_2|0,0^\mu\rangle_c \cr
&=&{}_c\langle0,0^\mu|\J_{2}\J_{-2}|0,0^\mu\rangle_c+{}_c\langle0,0^\mu|\bJ_{2}\bJ_{-2}|0,0^\mu\rangle_c \cr
=A_M(2)&=&\frac{1}{4}{}_c\langle0,0^\mu|C_1\cdot C_1 C_{-1}\cdot C_{-1}|0,0^\mu\rangle_c \cr
&&\quad+\frac{1}{4}{}_c\langle0,0^\mu|\C_1\cdot \C_1 \C_{-1}\cdot \C_{-1}|0,0^\mu\rangle_c=D.
\eea
From this we read the charges as $c_L=0$ and $c_M=2D$ which gives us the algebra \refb{BMS3} 
\bea{}
 [L_m,L_n]&=&(m-n)L_{m+n}+0, \cr
 [L_m,M_n]&=&(m-n)M_{m+n}+ \frac{D}{6}m(m^2-1)\delta_{m+n}, \cr
 [M_m,M_n]&=&0.
 \eea
 
 \subsection{Tensionless induced vacuum}\label{A3}

As seen above we need to consider ${}_I\langle0,0^\mu|[L_2,L_{-2}]|0,0^\mu\rangle_I$ and ${}_I\langle0,0^\mu|[L_2,M_{-2}]|0,0^\mu\rangle_I$. The action of $L_n$ on the vacuum can be read from (\ref{Linduced}) 
\be{} L_n|0,k^\mu\>_I=c'k\cdot A_n|0,k^\mu\>_I. \ee
For the zero momentum ground state we get 
\be{} L_n|0,k^\mu\>_I=0\quad \forall\ n \ee
Now we can apply this to the following sandwich and get for $m=1,2$
\be{} {}_I\<0,0^\mu|[L_m,L_{-m}]|0,0^\mu\>_I=A_L(m)=0. \ee
which gives us $c_L=0$. From \refb{Minduced} we get the action of $M_n$ on $|0,k^\mu\>_I$ which is simply
\be{} M_n|0,k^\mu\>_I=0\quad \forall\ n. \ee
Then the other sandwich can be calculated for $m=1,2$
\be{} {}_I\<0,0^\mu|[L_m,M_{-m}]|0,0^\mu\>_I= A_M(m)=0. \ee
from which we can read off $c_M=0$. Under the tensionless limt the tensile vacuum maps to the induced vacuum : $|0,0^\mu\>_\a\rightarrow|0,0^\mu\>_I$, hence the central charges follow from the limit. Since we have $c=\bar{c}$, this also implies 
\be{} c_L=c-\bar{c}=0\quad and\quad  c_M=\epsilon(c+\bar{c})\approx 0. \ee

\subsection{Tensionless flipped vacuum}\label{A2}
For the flipped vacuum we can write down the following relations by referring to \refb{f4} and \refb{f7} 
\bes\bea{}
L_n|0,0^\mu\>_A&=&0\quad\forall\ n\geq 0, \\
M_n|0,0^\mu\>_A&=&0\quad\forall\ n\geq 0. 
\eea\ees
Employing the oscillator structure from \refb{f3a} we can also write 
\bes\bea{}
L_{-1}|0,0^\mu\>_A&=&0;\quad L_{-2}|0,0^\mu\>_A=\frac{1}{2}\big[C_{-1}\cdot C_{-1}-\cc_{-1}\cdot \cc_{-1}\Big]|0,0^\mu\>_A, \\
M_{-1}|0,0^\mu\>_A&=&0; \quad M_{-2}|0,0^\mu\>_A=\frac{1}{2}\big[C_{-1}\cdot C_{-1}+\cc_{-1}\cdot \cc_{-1}\Big]|0,0^\mu\>_A. 
\eea\ees
Let us proceed to calculate the central charges. 
\bea{}
{}_A\langle0,0^\mu|[L_2,L_{-2}]|0,0^\mu\rangle_A&=&\frac{1}{4}{}_A\langle0,0^\mu|C_1\cdot C_1 C_{-1}\cdot C_{-1}|0,0^\mu\rangle_A \cr
&&\quad+\frac{1}{4}{}_A\langle0,0^\mu|\cc_1\cdot \cc_1 \cc_{-1}\cdot \cc_{-1}|0,0^\mu\rangle_A=D, \\
{}_A\langle0,0^\mu|[L_2,M_{-2}]|0,0^\mu\rangle_A&=&\frac{1}{4}{}_A\langle0,0^\mu|C_1\cdot C_1 C_{-1}\cdot C_{-1}|0,0^\mu\rangle_A \cr
&&\quad-\frac{1}{4}{}_A\langle0,0^\mu|\cc_1\cdot \cc_1 \cc_{-1}\cdot \cc_{-1}|0,0^\mu\rangle_A=0.
\eea
Collecting the above results we have $c_L=2A_L(2)=2D$ and $c_M=2A_M(2)=0$. This gives the centrally extended algebra as
\bea{}
[L_m, L_n] &=& (m-n) L_{m+n} + \frac{D}{6} m(m^2-1)\delta_{m+n,0}, \cr
[L_m, M_n] &=& (m-n) M_{m+n} + 0, \cr
[M_m, M_n] &=& 0.
\eea
This is consistent with the fact that due to flipping we have a change in the sign for the Virasoro central charge $\bar{c}\rightarrow-\bar{c}$. Then in the limit we will obtain $c_L=c+\bar{c}=2D$, while $c_M=\epsilon(c-\bar{c})=0$, which is precisely what is obtained above.

\newpage

\bibliographystyle{JHEP}
\bibliography{ref}

\end{document}